\documentclass[10pt,twocolumn,letterpaper]{article}

\usepackage{iccv}
\usepackage{times}
\usepackage{epsfig}
\usepackage{graphicx}
\usepackage{amsmath}
\usepackage{amssymb}

\usepackage{multirow}
\usepackage{subcaption}
\usepackage{tabularx}
\usepackage{longtable}
\usepackage{booktabs}
\usepackage[accsupp]{axessibility}

\usepackage[pagebackref=true,breaklinks=true,letterpaper=true,colorlinks,bookmarks=false]{hyperref}

\usepackage{color}

\definecolor{Purple}{rgb}{.8, 0, .8}
\definecolor{Red}{rgb}{1., 0, 0}
\definecolor{Blue}{rgb}{0, 0, 1.}
\definecolor{Green}{rgb}{0., .6, 0.}
\definecolor{Yellow}{rgb}{.9, .7, 0.}
\definecolor{Orange}{rgb}{1, 0.5, 0.}
\definecolor{Black}{rgb}{0., 0., 0.}
\newcommand{\ning}[1]{\textcolor{Black}{#1}}
\newcommand{\sahar}[1]{\textcolor{Black}{#1}}

\newcommand*{\affaddr}[1]{#1}
\newcommand*{\affmark}[1][*]{\textsuperscript{#1}}

\iccvfinalcopy 

\def\iccvPaperID{8062} 

\ificcvfinal\fi

\begin{document}

\title{\ning{Artificial Fingerprinting for Generative Models:\\Rooting Deepfake Attribution in Training Data}}

\author{Ning Yu\affmark[1,2]\thanks{Equal contribution.}\hspace{0.5cm}
Vladislav Skripniuk\affmark[3]$^*$\hspace{0.5cm}
Sahar Abdelnabi\affmark[3]\hspace{0.5cm}
Mario Fritz\affmark[3]\\
\affaddr{\affmark[1]University of Maryland}\hspace{0.5cm}
\affaddr{\affmark[2]Max Planck Institute for Informatics}\\
\affaddr{\affmark[3]CISPA Helmholtz Center for Information Security}\\
\tt\small \{ningyu,vladislav\}@mpi-inf.mpg.de\hspace{0.5cm}
\tt\small \{sahar.abdelnabi,fritz\}@cispa.saarland
}

\maketitle
\ificcvfinal\thispagestyle{empty}\fi

\begin{abstract}
Photorealistic image generation has reached a new level of quality due to the breakthroughs of generative adversarial networks (GANs). Yet, the dark side of such deepfakes, the malicious use of generated media, raises concerns about visual misinformation. While existing research work on deepfake detection demonstrates high accuracy, it is subject to advances in generation techniques and adversarial iterations on detection countermeasure techniques. Thus, we seek a proactive and sustainable solution on deepfake detection, that is agnostic to the evolution of generative models, by introducing artificial fingerprints into \ning{the models}. 

\ning{Our approach is simple and effective.} We first embed artificial fingerprints into training data, then validate a surprising discovery on the transferability of such fingerprints from training data to generative models, which in turn \ning{appears in the generated deepfakes}. Experiments show that our fingerprinting solution (1) holds for a variety of cutting-edge \ning{generative models}, (2) leads to a negligible side effect on generation quality, (3) stays robust against image-level and model-level perturbations, (4) stays hard to be detected by adversaries, and (5) \ning{converts deepfake detection and attribution into trivial tasks and outperforms the recent state-of-the-art baselines. Our solution closes the responsibility loop between \sahar{publishing pre-trained} generative model inventions and their possible misuses}, which makes it independent of the current arms race. Code and models are available at \href{https://github.com/ningyu1991/ArtificialGANFingerprints}{GitHub}.
\end{abstract}

\section{Introduction}
In the past years, photorealistic image generation has been rapidly evolving, benefiting from the invention of generative adversarial networks (GANs)~\cite{gan14nips} and its successive breakthroughs~\cite{radford2015dcgan,gulrajani2017wgan,miyato2018spectral,brock2018BigGAN,karras2017ProGAN,karras2019StyleGAN,karras2019StyleGAN2,yu2020inclusive,yu2021dual}. Given the level of realism and diversity that generative models can achieve today, detecting generated media, well known as \textit{deepfakes}, attributing their sources, and tracing their legal responsibilities become infeasible to human beings.

Moreover, the misuse of deepfakes has been permeating to each corner of social media, ranging from misinformation of political campaigns~\cite{misinformationPoliticalCampaigns} to fake journalism~\cite{fakejournalism1,fakejournalism2}. This motivates tremendous research efforts on deepfake detection~\cite{zhang2020not} and source attribution~\cite{marra2019gans,ning2019iccv_gan_detection,wang2020cnn}. These techniques aim to counter the widespread of malicious applications of deepfakes by automatically identifying and flagging generated visual contents and tracking their sources. Most of them rely on low-level visual patterns in GAN-generated images~\cite{marra2019gans,ning2019iccv_gan_detection,wang2020cnn,he2021beyond,zhou2021deep} or frequency mismatch~\cite{durall2019unmasking,zhang2019detecting,frank2020dct2d_detect}. However, these techniques are unable to sustainably and robustly prevent deepfake misuse in the long run; as generative models evolve, they learn to better match the true distribution causing fewer artifacts~\cite{zhang2020not}. Besides, detection countermeasures are also continuously evolving~\cite{margret2020upconvolution,carlini2020evading,zhang2020not}.

\begin{figure*}[!t]
\centering
\includegraphics[width=\linewidth]{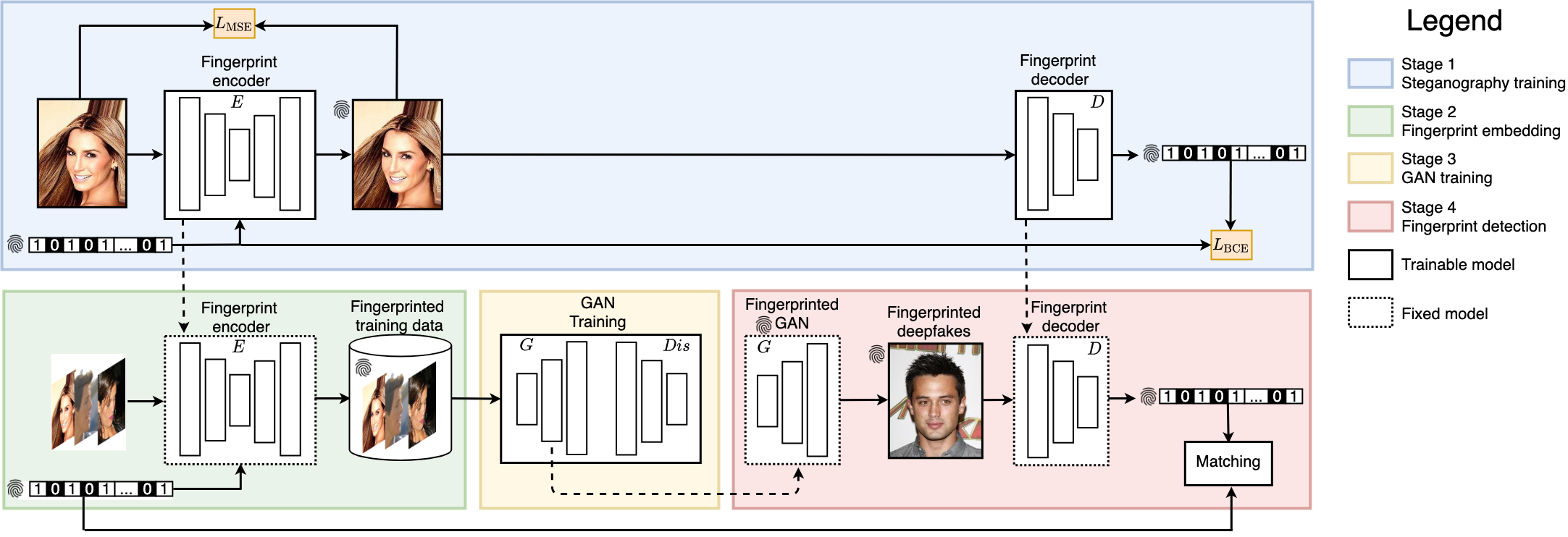}
\caption{Our solution pipeline consists of four stages. We first train an image steganography encoder and decoder. Then we use the encoder to embed artificial fingerprints into the training data. After that, we train a generative model with its original protocol. Finally, we decode the fingerprints from the generated deepfakes.}
\label{fig:teaser}
\vspace{-4mm}
\end{figure*}

Motivated by this, we tackle deepfake detection and attribution through a different lens, and propose a \ning{\textbf{\textit{proactive}}} and sustainable solution for detection\ning{, which is simple and effective.} In specific, we aim to introduce \textbf{\textit{artificial fingerprints}} into \ning{generative models} that enable identification and tracing. Figure~\ref{fig:teaser} depicts our pipeline; we first embed artificial fingerprints into the training data using image steganography~\cite{baluja2017hiding,tancik2019stegastamp}. The generative model is then trained with its original protocol without modification. \ning{This makes our solution agnostic and plug-and-play for arbitrary models.} We then show a surprising discovery on the \textbf{\textit{transferability}} of such fingerprints from training data to the model: the same fingerprint information that was encoded in the training data can be decoded from all generated images. 

We achieve deepfake detection by classifying images with matched fingerprints in our database as fake and images with random detected fingerprints as real. 
We also achieve deepfake attribution when we allocate different fingerprints for different generative models. Our solution thus \ning{closes the responsibility loop between generative model inventions and their possible misuses.} It prevents the misuse of \sahar{published pre-trained} generative models by enabling inventors to proactively and responsibly embed artificial fingerprints into the models.

We summarize our contributions as follow:

(1) \ning{We synergize the two previously uncorrelated domains, image steganography and GANs, and propose the first \textbf{\textit{proactive}} and sustainable solution for the third emerging domain, deepfake detection and attribution.}

(2) This is the first study to demonstrate the \textbf{\textit{transferability}} of artificial fingerprints from training data to generative models and then to all the generated deepfakes. 
\ning{Our discovery is non-trivial: only deep-learning-based fingerprinting techniques~\cite{baluja2017hiding,tancik2019stegastamp} are transferable to generative models, while conventional steganography and watermarking techniques~\cite{outguess,steghide} are not. See Section~\ref{sec:effectiveness} for comparisons.}

(3) We empirically validate several beneficial properties of our solution. \textbf{\textit{Universality}} (Section~\ref{sec:effectiveness}): it holds for a variety of cutting-edge generative models~\cite{karras2017ProGAN,karras2019StyleGAN,karras2019StyleGAN2,brock2018BigGAN,park2020cut}. \textbf{\textit{Fidelity}} (Section~\ref{sec:fidelity}): it has a negligible side effect on generation quality. \textbf{\textit{Robustness}} (Section~\ref{sec:robustness}): it stays robust against many perturbations. \textbf{\textit{Secrecy}} (Section~\ref{sec:secrecy}): the artificial fingerprints are hard to be detected by adversaries. \ning{\textbf{\textit{Anti-deepfake}} (Section~\ref{sec:detection} and \ref{sec:attribution}): it converts deepfake detection and attribution into trivial tasks and outperforms the state-of-the-art baselines~\cite{ning2019iccv_gan_detection,wang2020cnn}.}

\section{Related Work}
\textbf{Generative adversarial networks (GANs).} GANs~\cite{gan14nips} was first proposed as a workaround to model the intractable real data distribution. The iterative improvements push the generation realism to brand-new levels~\cite{radford2015dcgan,gulrajani2017wgan,miyato2018spectral,brock2018BigGAN,karras2017ProGAN,karras2019StyleGAN,karras2019StyleGAN2}. Successes have also spread to many other vision tasks (e.g.~\cite{park2019semantic,ledig2017photo,isola2017image,zhu2017unpaired,zhu2017toward,park2020cut,yu2018generative}). 
\ning{In Section~\ref{sec:exp}, we focus on three categories of cutting-edge generative models:  unconditional (ProGAN~\cite{karras2017ProGAN}, StyleGAN~\cite{karras2019StyleGAN}, and StyleGAN2~\cite{karras2019StyleGAN2}), class-conditional (BigGAN~\cite{brock2018BigGAN}), and image-conditional (image-to-image translation) (CUT~\cite{park2020cut}).}

\textbf{Image steganography and watermarking.} Image steganography and watermarking hide information into carrier images~\cite{fridrich2009steganography}. Previous techniques rely on Fourier transform~\cite{cox2002digital,cayre2005watermarking}, JPEG compression~\cite{outguess,steghide}, or least significant bits modification~\cite{pevny2010using,holub2012designing,holub2014universal}. Recent works replace hand-crafted hiding procedures with neural network encoding~\cite{baluja2017hiding,hayes2017generating,vukotic2018deep,zhu2018hidden,zhang2019invisible,tancik2019stegastamp,luo2020distortion}. We leverage recent deep-learning-based steganography methods~\cite{baluja2017hiding,tancik2019stegastamp} to embed artificial fingerprints into training data, and validate their transferability to generative models. \ning{This is non-trivial because only deep-learning-based fingerprints are transferable to generative models, while conventional ones~\cite{outguess,steghide} are not (Section~\ref{sec:effectiveness}). Besides,} the stealthiness achieved by steganography allows preserving the original generation quality (Section~\ref{sec:fidelity}) and fingerprint secrecy (Section~\ref{sec:secrecy}).

\ning{Our fingerprinting is conceptually and functionally orthogonal to all of them. Instead of encoding information into pixels of individual images, our solution encodes information into generator parameters such that all the generated images are entangled with that information. Compared to the pipeline of a generator followed by a watermarking module,  our solution introduces zero generation overheads, and obstructs adversarial model surgery that targets to detach watermarking from image generation.}

\textbf{Network watermarking.} Different from image watermarking, network watermarking targets to hide information into model parameters without affecting its original performances, similar in spirit to our goal. There are two categories of them: black-box trigger-set-based solutions~\cite{adi2018turning,zhang2018protecting}, and white-box feature-based solutions~\cite{uchida2017embedding,chen2019deepmarks,rouhani2019deepsigns}. The former ones embed watermarks through a trigger set of input and decodes watermarks according to the input-output behavior of the model. The latter ones directly embed watermarks in the model parameter space with transformation matrices. It is worth noting that our solution renders conceptual and technical distinctions from network watermarking. In terms of concepts, the previous works target to only discriminative models (e.g., classification), while a solution for generative models is urgently lacking. In terms of techniques, to adapt to generator watermarking, we tune our solution to \textit{indirectly} transfers fingerprints from training data to model parameters. This is because (1) unconditional generative models do not allow deterministic input so that a trigger set is not applicable, and (2) transformations in the parameter space are not agnostic to model configurations so that they are neither scalable nor sustainable along with the evolution of generative models.

\textbf{Deepfake detection and attribution.} Images generated by GAN models bear unique patterns. \cite{marra2019gans} shows that generative models leave unique noise residuals to generated samples, which allows deepfake detection. \cite{ning2019iccv_gan_detection} moves one step further, using a neural network classifier to attribute different images to their sources. \cite{wang2020cnn} also train a classifier and improve the generalization across different generation techniques. \cite{zhang2019detecting,durall2019unmasking,margret2020upconvolution} point out that the high-frequency pattern mismatch can 
be used for deepfake detection, so can the texture feature mismatch~\cite{liu2020texture_fake}. However, these cues are not sustainable 
due to the advancement of detection countermeasures. For example, spectral regularization~\cite{margret2020upconvolution} is proposed to narrow down the frequency mismatch and results in a significant detection deterioration. Also, detectors~\cite{wang2020cnn} are vulnerable to adversarial evasion attacks~\cite{carlini2020evading}. 

In contrast to the previous passive approaches, we propose a novel \textit{proactive} solution for model fingerprinting and, thus, for deepfake detection. We differentiate between our term \textit{artificial fingerprints} which refers to the information we deliberately and proactively embed into the model, and the term GAN fingerprints~\cite{ning2019iccv_gan_detection} which refers to the inherent cues and artifacts of different GAN models. \ning{Our work is also distinct from a follow-up proactive technique~\cite{yu2020responsible}. They focus on fingerprinting scalability and efficiency while we focus more fundamentally on its transferability and universality.}

\section{Problem Statement}
Generation techniques can be misused to create misinformation at scale 
to achieve financial or political gains. Recently, there have been concerns about releasing generative models. For example, OpenAI 
employed a staged release to evaluate the potential risks of their \sahar{GPT-2} model~\cite{radford2019language}. \sahar{GPT-3 was later released as a black-box API only~\cite{brown2020language}.} \ning{Face2Face~\cite{thies2016face2face} authors did not open their sources for real-time face capture and reenactment.}

We design solution from the model inventors' side (e.g., OpenAI). Our solution introduces traceable artificial fingerprints in generative models. It enables deepfake detection and attribution by decoding the fingerprints from the generated images and matching them to the known fingerprints given to different models.
This equips model inventors with a means for a proactive and responsible disclosure when publishing their \sahar{pre-trained }models. This distinguishes our model fingerprinting solution from watermarking the generated images: we aim to defend against the misuse of published generative models rather than single deepfake media. 

In practice, the training is done by the model inventor. Responsible model inventors, different from malicious deepfake users, should be \textit{eager}/\textit{willing} to adopt a proactive solution to fingerprint their generative models against potential deepfake misuses. \ning{The fingerprinting encoder and decoder, and the unique fingerprints given to different models, are privately maintained by the model inventor. Once a deepfake misuse happens, the inventor is able to verify if this is generated by one of their models. If so, they can further attribute by which model user. 
Then they can prohibit that user's accessibility to the model and/or seek legal regulations. Thus, they can claim responsible disclosure with a countermeasure against potential misuse when they publish their models.}

\section{Artificial Fingerprints}
\label{sec:method}
The goal of image attribution is to learn a mapping $D_0(\mathbf{x}) \mapsto y$ that traces the source $y \in \mathbb{Y} = \{\text{real}, \text{G}_\text{1}, \dots, \text{G}_\text{N}\}$ of an image $\mathbf{x}$. If the domain $\mathbb{Y}$ is limited, predefined, and known to us, this is a closed-world scenario and the attribution can be simply formulated as a multi-label classification problem, each label corresponding to one source, as conducted in~\cite{ning2019iccv_gan_detection}. However, $\mathbb{Y}$ can be unlimited, undefined, continuously evolving, and agnostic to us. This open-world scenario is intractable using discriminative learning. To generalize our solution to being agnostic to the selection of generative models, we formulate the attribution as a regression mapping $D(\mathbf{x}) \mapsto \mathbf{w}$, where $\mathbf{w} \in \{0,1\}^n$ is the source identity space and $n$ is the dimension. 
We propose a pipeline to root the attribution down to the training dataset $\tilde{\mathbf{x}} \in \tilde{\mathbb{X}}$ and close the loop of the regression $D$. We describe the pipeline stages (depicted in Figure~\ref{fig:teaser}) below: 

\textbf{Steganography training.}
The source identity is represented by the artificial fingerprints $\mathbf{w}$. We use a steganography system~\cite{baluja2017hiding,tancik2019stegastamp} to learn an encoder $E(\tilde{\mathbf{x}}, \mathbf{w}) \mapsto \tilde{\mathbf{x}}_\mathbf{w}$ that embeds an arbitrary fingerprint $\mathbf{w}$ (randomly sampled during training) into an arbitrary image $\tilde{\mathbf{x}}$. We couple $E$ with a decoder $D(\tilde{\mathbf{x}}_\mathbf{w}) \mapsto \mathbf{w}$ to detect the fingerprint information from the image. $E$ and $D$ are formulated as convolutional neural networks with the following training losses:
\begin{equation} 
    \min_{E,D} \mathbb{E}_{\tilde{\mathbf{x}}\sim\tilde{\mathbb{X}},\mathbf{w}\sim\{0,1\}^n} L_\text{BCE}(\tilde{\mathbf{x}},\mathbf{w};E,D) +
\lambda L_\text{MSE}(\tilde{\mathbf{x}},\mathbf{w};E)
\label{eq:objective}
\end{equation}
\begin{equation}
L_\text{BCE}(\tilde{\mathbf{x}},\mathbf{w};E,D) = \frac{1}{n}\sum_{k=1}^n \big( \mathbf{w}_k \log \hat{\mathbf{w}}_k +(1 - \mathbf{w}_k) \log(1 - \hat{\mathbf{w}}_k)\big)
\label{eq:loss_BCE}
\end{equation}
\begin{equation}
L_\text{MSE}(\tilde{\mathbf{x}},\mathbf{w};E) = ||E(\tilde{\mathbf{x}},\mathbf{w}) - \tilde{\mathbf{x}}||_2^2
\end{equation}
\begin{equation}
\hat{\mathbf{w}} = D\big(E(\tilde{\mathbf{x}},\mathbf{w})\big)
\end{equation}
where $\mathbf{w}_k$ and $\hat{\mathbf{w}}_k$ are the $k^{\text{th}}$ bit of the input fingerprint and detected fingerprint separately; and $\lambda$ is a hyper-parameter to balance the two objective terms. The binary cross-entropy term $L_\text{BCE}$ guides the decoder to decode the fingerprint embedded by the encoder. The mean squared error term $L_\text{MSE}$ penalizes any deviation of the stego image $E(\tilde{\mathbf{x}},\mathbf{w})$ from the original image $\tilde{\mathbf{x}}$. The architectures of $E$ and $D$ are depicted in the supplementary material.

\textbf{Artificial fingerprint embedding.} In this stage, we use the well trained $E$ and $D$ networks. We allocate each training dataset $\tilde{\mathbb{X}}$ a unique fingerprint $\mathbf{w}$. We apply the trained $E$ to each training image $\tilde{\mathbf{x}}$ and collect a fingerprinted training dataset $\tilde{\mathbb{X}}_\mathbf{w} = \{E(\tilde{\mathbf{x}},\mathbf{w}) | \tilde{\mathbf{x}} \in \tilde{\mathbb{X}}\}$.

\textbf{Generative model training.} 
\ning{In order to have a solution that is agnostic to the evolution of generative models, we intentionally do not intervene with their training. It makes our solution plug-and-play for arbitrary generation tasks without touching their implementations, and introduces zero overhead to model training.} We simply replace $\tilde{\mathbb{X}}$ with $\tilde{\mathbb{X}}_\mathbf{w}$ to train the generative model in its original protocol. 

\textbf{Artificial fingerprint decoding.} We hypothesize the \textit{transferability} of our artificial fingerprints from training data to generative models: a well-trained generator $G_\mathbf{w}(\mathbf{z}) \mapsto \mathbf{x}_\mathbf{w}$ contains, in all generated images, the same fingerprint information $\mathbf{w}$ (as embedded in the training data $\tilde{\mathbf{x}}_\mathbf{w}$). \ning{We justify this hypothesis in Section~\ref{sec:effectiveness}}. As a result, the artificial fingerprint can be recovered from a generated image $\mathbf{x}_\mathbf{w}$ using the decoder $D$: $D(\mathbf{x}_\mathbf{w}) \equiv \mathbf{w}$. Based on this transferability, we can formulate deepfake attribution as fingerprint matching using our decoder $D$.

\textbf{Artificial fingerprint matching.} To support robustness to post-generation modifications that could be applied to the generated images, we relax the matching of the decoded artificial fingerprints to a soft matching. We perform a null hypothesis test given the number of matching bits $k$ between the decoded fingerprint $\tilde{\mathbf{w}}$ and the fingerprint $\mathbf{w}$ used in generative model training. The null hypothesis $H_0$ is getting this number of successes (i.e. matching bits) by chance. Under the null hypothesis, the probability of matching bits (random variable $X$) follows a binomial distribution: the number of trials $n$ is the number of bits in the fingerprint sequence, and $k$ is the number of successes where each bit has a 0.5 probability of success. We can then measure the $p$-value of the hypothesis test by computing the probability of getting $k$ or higher matching bits under the null hypothesis: 
\begin{equation}
Pr(X>k|H_0) = \sum_{i=k}^{n} \binom{n}{i} 0.5^n 
\end{equation}
The fingerprint is verified, $\tilde{\mathbf{w}} \sim \mathbf{w}$, if the null hypothesis results in a very low probability ($p$-value). 
Usually, when the $p$-value is smaller than $0.05$, we reject the null hypothesis and regard $1-p$ as the verification confidence.

\section{Experiments}
\label{sec:exp}
We describe the experimental setup in Section~\ref{sec:setup}. We first evaluate the required proprieties of our solution: the transferability and universality of our artificial fingerprint in Section~\ref{sec:effectiveness}, its fidelity in Section~\ref{sec:fidelity}, its robustness in Section~\ref{sec:robustness}, 
and its secrecy in Section~\ref{sec:secrecy}. 
The transferability in turn enables accurate deepfake detection and attribution, which is evaluated and compared in Section~\ref{sec:detection} and \ref{sec:attribution} respectively. 
In addition, we articulate our network designs and training details in the supplementary material.

\subsection{Setup}
\label{sec:setup}

\textbf{Generative models.} \ning{As a proactive solution, it should be agnostic to genetative models. Without losing representativeness, we focus on three generation applications with their state-of-the-art models. For unconditional generation: ProGAN~\cite{karras2017ProGAN}, StyleGAN~\cite{karras2019StyleGAN}, and StyleGAN2~\cite{karras2019StyleGAN2}; for class-conditional generation: BigGAN~\cite{brock2018BigGAN}; for image-conditional generation, i.e., image-to-image translation: CUT~\cite{park2020cut}. Each model is trained from scratch with the official implementation.}

\textbf{Datasets.} \ning{Each generation application benchmarks its own datasets. For unconditional generation, we train/test on 150k/50k CelebA~\cite{liu2015faceattributes} at 128$\times$128 resolution, 50k/50k LSUN \textit{Bedroom}~\cite{yu2015lsun} at 128$\times$128 resolution, and the most challenging one, 50k/50k LSUN \textit{Cat}~\cite{yu2015lsun} at its original 256$\times$256 resolution. For class-conditional generation, we experiment on the entire CIFAR-10 dataset~\cite{krizhevsky2009learning} with the original training/testing split at the original 32$\times$32 resolution. For image-conditional generation, we experiment on the entire \textit{Horse}\textrightarrow\textit{Zebra} dataset~\cite{zhu2017unpaired} and \textit{Cat}\textrightarrow\textit{Dog}~\cite{choi2020stargan} dataset with the original training/testing split at the original 256$\times$256 resolution. We only need to fingerprint images from the target domains.}

\subsection{Transferability}
\label{sec:effectiveness}

The transferability means that the artificial fingerprints that are embedded in the training data also appear consistently in all the generated data. This is a non-trivial hypothesis in Section~\ref{sec:method} and needs to be justified by the fingerprint detection accuracy.

\textbf{Evaluation.} Fingerprints are represented as binary vectors $\mathbf{w} \in \{0,1\}^n$. We use bitwise accuracy to evaluate the detection accuracy. We set $n = 100$ as suggested in~\cite{tancik2019stegastamp}. We also report $p$-value for the confidence of detection.

\textbf{Baselines.} For comparison, we implement a straightforward baseline method. Instead of embedding fingerprints into training data, we enforce fingerprint generation jointly with model training. That is, we train on clean data, and enforce generated images to not only approximate real training images but also contain a specific fingerprint. Mathematically,
\begin{equation} 
\begin{aligned}
\min_{G,D}\max_{Dis} \mathbb{E}_{\mathbf{z}\sim\mathcal{N}(\mathbf{0},\mathbf{I}),\tilde{\mathbf{x}}\sim\tilde{\mathbb{X}}}L_\text{adv}(\mathbf{z},\tilde{\mathbf{x}};G,Dis) +\\
\eta\mathbb{E}_{\mathbf{z}\sim\mathcal{N}(\mathbf{0},\mathbf{I}),\mathbf{w}\sim\{0,1\}^n}L_\text{BCE}(\mathbf{z},\mathbf{w};G,D)
\end{aligned}
\label{eq:baseline}
\end{equation}
where $G$ and $Dis$ are the original generator and discriminator in the GAN framework, $L_\text{adv}$ is the original GAN objective, and $L_\text{BCE}$ is adapted from Eq.~\ref{eq:loss_BCE} where we replace $\hat{\mathbf{w}} = D(E(\tilde{\mathbf{x}},\mathbf{w}))$ with $\hat{\mathbf{w}} = D(G(\mathbf{z}))$. $\eta$ is set to 1.0 as a hyper-parameter to balance the two objective terms.

\sahar{We also compare the deep-learning-based steganography technique used in our solution (\cite{tancik2019stegastamp}) to two well-established, non-deep learning steganographic methods~\cite{outguess,steghide} that alter the frequency coefficients of JPEG compression.}

\textbf{Results.} We report the fingerprint detection performance in Table~\ref{tab:accuracy_fid} fourth and fifth columns. We observe:

(1) The ``Data'' row shows the detection accuracy on real testing images for sanity checks: it reaches the 100\% saturated accuracy, indicating the effectiveness of the steganography technique by its nature.

(2) Our artificial fingerprints can be almost perfectly and confidently detected from generated images over a variety of applications, generative models, and datasets. The accuracy is $\geq 0.98$ except for ProGAN on LSUN \textit{Bedroom}, but its $0.93$ accuracy and $10^{-19}$ p-value are far sufficient to verify the presence of fingerprints. Our hypothesis on the \textit{transferability} from training data to generative models (i.e. generated data) is therefore justified. As a result, artificial fingerprints are qualified for deepfake detection and attribution. 

(3) The \textit{universality} of fingerprint transferability over varying tasks and models validates our solution is agnostic to generative model techniques. 

(4) \sahar{The baseline of joint fingerprinting and generation training} (first row) is \ning{also moderately effective in terms of fingerprint detection, but we show in Section~\ref{sec:fidelity} it leads to strong deterioration of generation quality.}

(5) \sahar{Conventional steganography methods~\cite{outguess,steghide} (second and third rows) do not transfer hidden information into models, indicated by the random guess performance during decoding. We attribute this to the discrepancy between deep generation techniques and shallow steganography techniques. We reason that generative models leverage deep discriminators to approximate common image patterns including low-level fingerprints. Only comparably deep-learning-based fingerprinting techniques, e.g.~\cite{tancik2019stegastamp}, are compatible to hide and transfer fingerprints to the models, while hand-crafted image processing is not effective.} \ning{Therefore, the transferability of our fingerprinting is non-trivial.}

\begin{table}[t!]
\center
\resizebox{\linewidth}{!}{%
\begin{tabular}{lll|cc|cc}\toprule
 & Fgpt & & Bit & & Orig & Fgpt \\ 
Dataset & tech & Model & acc $\Uparrow$ & $p$-value & FID & FID $\Downarrow$\\ 
\midrule
\multirow{7}{*}{CelebA} & Eq.~\ref{eq:baseline} & ProGAN &  0.93 & $<10^{-19}$ & 14.09 & 60.28 \\ 
 & \cite{outguess} & StyleGAN2 & 0.51 & 0.46 & 6.41 & 6.93 \\
 & \cite{steghide} & StyleGAN2 & 0.53 & 0.31 & 6.41 & 6.82 \\ \cmidrule{2-7}
 & \cite{tancik2019stegastamp} & Data & 1.00 & - & - & 1.15\\
 & \cite{tancik2019stegastamp} & ProGAN & 0.98 & $<10^{-26}$ & 14.09 & 14.38 \\
 & \cite{tancik2019stegastamp} & StyleGAN & 0.99 & $<10^{-28}$ & 8.98 & 9.72 \\
 & \cite{tancik2019stegastamp} & StyleGAN2 & 0.99 & $<10^{-28}$ & 6.41 & 6.23 \\
\midrule
\multirow{3}{*}{LSUN} 
 & \cite{tancik2019stegastamp} & ProGAN &  0.93 & $<10^{-19}$ & 29.16 & 32.58 \\
 & \cite{tancik2019stegastamp} & StyleGAN & 0.98 & $<10^{-26}$ & 24.95 & 25.71\\
\textit{Bedroom} & \cite{tancik2019stegastamp} & StyleGAN2 & 0.99 & $<10^{-28}$ & 13.92 & 14.71 \\
\midrule
\multirow{3}{*}{LSUN} & \cite{tancik2019stegastamp} & ProGAN & 0.98 & $<10^{-26}$ & 45.22 & 48.97 \\
 & \cite{tancik2019stegastamp} & StyleGAN & 0.99 & $<10^{-28}$ & 33.45 & 34.01 \\
\textit{Cat} & \cite{tancik2019stegastamp} & StyleGAN2 & 0.99 & $<10^{-28}$ & 31.01 & 32.60 \\
\midrule
CIFAR-10 & \cite{tancik2019stegastamp} & BigGAN & 0.99 & $<10^{-28}$& 6.25 & 6.80  \\
\midrule
\textit{Horse}$\rightarrow$\textit{Zebra} & \cite{tancik2019stegastamp} & CUT & 0.99 & $<10^{-28}$& 22.98 & 23.43   \\
\textit{Cat}$\rightarrow$\textit{Dog} & \cite{tancik2019stegastamp} & CUT & 0.99 & $<10^{-28}$& 55.78 & 56.09  \\
\bottomrule
\end{tabular}}
\caption{Artificial fingerprint detection in bitwise accuracy ($\Uparrow$ indicates higher is better) and generation quality in FID ($\Downarrow$ indicates lower is better). The ``Data'' row corresponds to real testing images for a sanity check. The ``Orig FID'' column corresponds to the original (non-fingerprinted) models for references. \sahar{The first three rows are the baselines.}}
\label{tab:accuracy_fid}
\end{table}

\begin{figure*}[t!]
\begin{subfigure}[b]{0.19\linewidth}
\centering
\includegraphics[width=\linewidth]{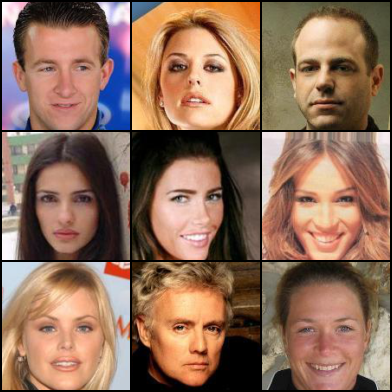}
\caption{}
\label{subfig:orig_train}
\end{subfigure} 
\begin{subfigure}[b]{0.19\linewidth}
\centering
\includegraphics[width=\linewidth]{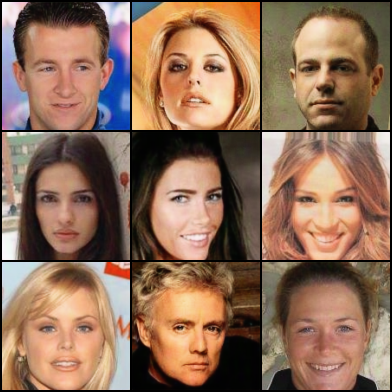} 
\caption{}
\label{subfig:watermarked_train}
\end{subfigure} 
\begin{subfigure}[b]{0.19\linewidth}
\centering
\includegraphics[width=\linewidth]{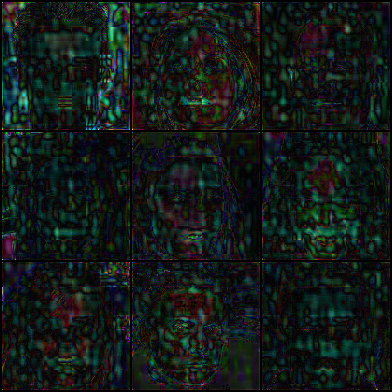}
\caption{}
\label{subfig:diff}
\end{subfigure}
\begin{subfigure}[b]{0.19\linewidth}
\centering
\includegraphics[width=\linewidth]{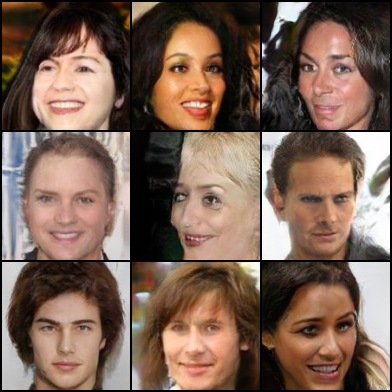}
\caption{}
\label{subfig:clean_fake}
\end{subfigure} 
\begin{subfigure}[b]{0.19\linewidth}
\centering
\includegraphics[width=\linewidth]{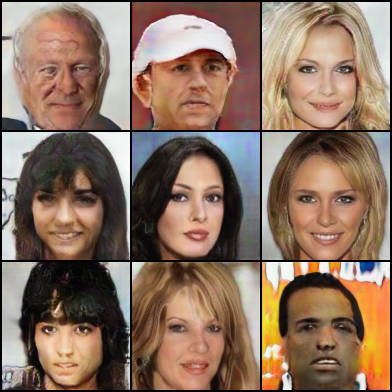}
\caption{}
\label{subfig:stego_fake}
\end{subfigure}
\caption{CelebA samples at 128$\times$128 for Table~\ref{tab:accuracy_fid} last two columns. (a) Original real training samples. (b) Fingerprinted real training samples. (c) The difference between (a) and (b), $10\times$ magnified for easier visualization. (d) Samples from the non-fingerprinted ProGAN. (e) Samples from the fingerprinted ProGAN. See more samples on the other datasets in the supplementary material.}
\label{fig:watermark_samples} 
\end{figure*}

\subsection{Fidelity}
\label{sec:fidelity}

The fidelity of generated images is as critical as the transferability. 
Fingerprinting should have a negligible side effect on the functionality of generative models. This preserves the original generation quality and avoids the adversary's suspect of the presence of fingerprints. The steganography technique we used should enable this, which we validate empirically.

\textbf{Evaluation.} We use Fr\'{e}chet Inception Distance (FID)~\cite{heusel2017gans} to evaluate the generation quality; the lower, the more realistic. We measure FID between a set of 50k generated images and a set of 50k real non-fingerprinted images, in order to evaluate the quality of the generated set. When calculating different FIDs for each dataset, the real set is unchanged.

\textbf{Results.} We compare the generation quality of original and fingerprinted generative models in Table~\ref{tab:accuracy_fid} sixth and seventh columns. We observe:

(1) The ``Data'' rows are for sanity checks: embedding fingerprints into real images does not substantially deteriorate image quality: FID $\leq 1.15$ is in an excellent realism range. This validates the secrecy of the steganographic technique and lays a valid foundation for high-quality model training.

(2) For a variety of settings, the performance of the fingerprinted generative models tightly sticks to the original limits of their non-fingerprinted baselines. \ning{The heaviest deterioration is as small as $+3.75$ FID happening for ProGAN on LSUN \textit{Cat}.} In practice, the generated fingerprints are imperceptibly hidden in the generated images and can only be perceived under $10\times$ magnification. See Figure~\ref{fig:watermark_samples} and the supplementary material for demonstrations. Therefore, the fidelity of fingerprinted models is justified and it qualifies our solution for deepfake detection and attribution. 

(3) \ning{The baseline of joint fingerprinting and generation training (first row) deteriorates generation quality remarkably. This indicates model fingerprinting is a non-trivial task: direct fingerprint reconstruction distracts adversarial training. In contrast, our solution leverages image steganography and fingerprint transferability, sidesteps this issue, and leads to better performance.}

\subsection{Robustness}
\label{sec:robustness}

Deepfake media and generative models may undergo post-processing or perturbations during broadcasts. We validate the robustness of our fingerprint detection given a variety of image and model perturbations, and investigate the corresponding working ranges.

\textbf{Perturbations.} We evaluate the robustness against four types of \textit{image perturbation}: additive Gaussian noise, blurring with Gaussian kernel, JPEG compression, center cropping. We also evaluate the robustness against two types of \textit{model perturbations}: model weight quantization and adding Gaussian noise to model weights. For quantization, we compress each model weight given a decimal precision. We vary the amount of perturbations, apply each to the generated images or to the model directly, and detect the fingerprint using the pre-trained decoder.

\begin{figure*}[t!]
  \begin{subfigure}[b]{0.16\linewidth}
    \centering
    \includegraphics[width=\linewidth]{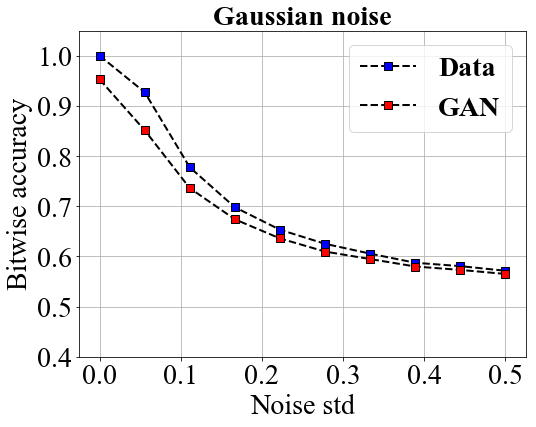}
  \end{subfigure}
  \begin{subfigure}[b]{0.16\linewidth}
    \centering
    \includegraphics[width=\linewidth]{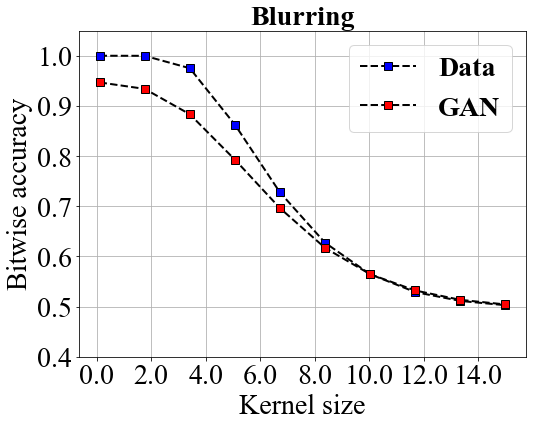} 
  \end{subfigure}
  \begin{subfigure}[b]{0.16\linewidth}
    \centering
    \includegraphics[width=\linewidth]{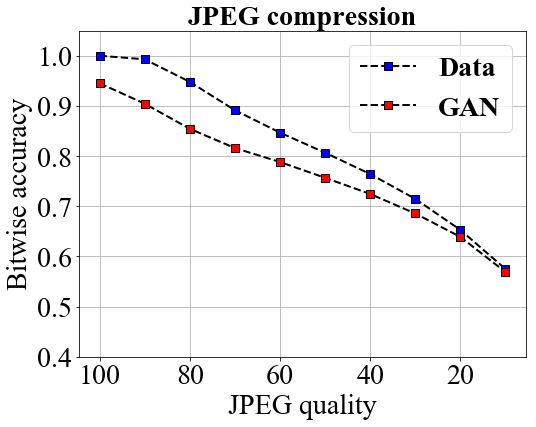} 
  \end{subfigure}
  \begin{subfigure}[b]{0.16\linewidth}
    \centering
    \includegraphics[width=\linewidth]{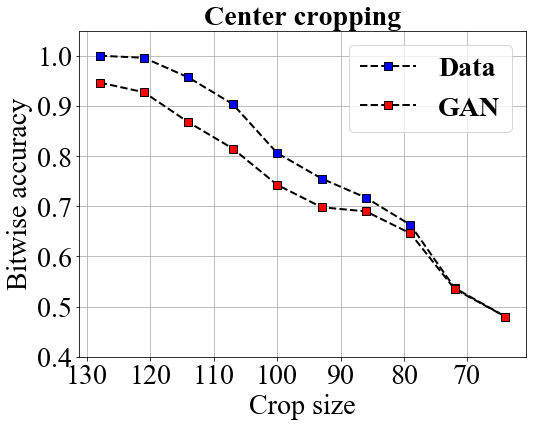} 
  \end{subfigure}
  \begin{subfigure}[b]{0.17\linewidth}
    \centering
    \includegraphics[width=\linewidth]{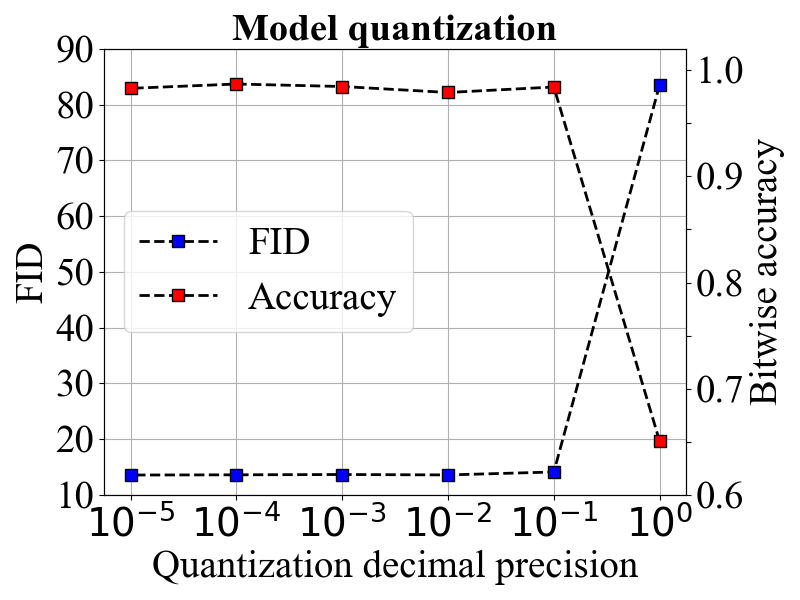} 
  \end{subfigure}
  \begin{subfigure}[b]{0.17\linewidth}
    \centering
    \includegraphics[width=\linewidth]{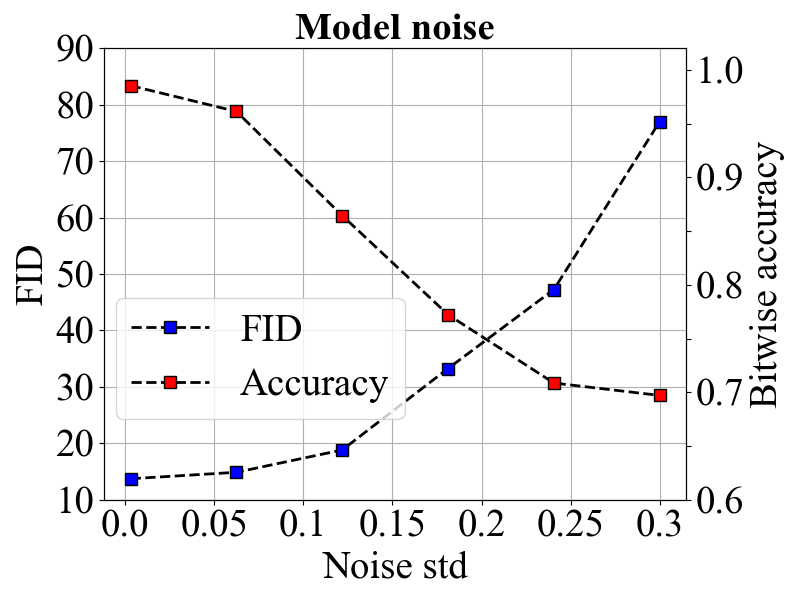} 
  \end{subfigure}
  \caption{Red plots show the artificial fingerprint detection in bitwise accuracy w.r.t. the amount of perturbations over ProGAN trained on CelebA. In the left four plots (robustness against \textit{image perturbations}), blue dots represent detection accuracy on the fingerprinted real training images, which serve as the upper bound references for the red dots. See the supplementary material for additional results of ProGAN trained on LSUN \textit{Bedroom}. In the right two plots (robustness against \textit{model perturbations}), blue dots represent the FID of generated images from the perturbed models.}
  \label{fig:robutstness_plots} 
\end{figure*}

\begin{figure*}[t!]
  \begin{subfigure}[b]{0.14\linewidth}
    \centering
    \includegraphics[width=\linewidth]{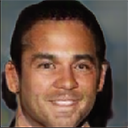}
    \captionsetup{justification=centering}
    \caption{\\Original\\0.99 bit acc}
  \end{subfigure}
  \begin{subfigure}[b]{0.14\linewidth}
    \centering
    \includegraphics[width=\linewidth]{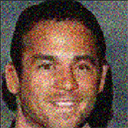}
    \captionsetup{justification=centering}
    \caption{Gaussian noise\\Std 0.1\\0.77 bit acc}
  \end{subfigure}
  \begin{subfigure}[b]{0.14\linewidth}
    \centering
    \includegraphics[width=\linewidth]{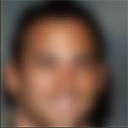}
    \captionsetup{justification=centering}
    \caption{Blurring\\Kernel size 5\\0.75 bit acc}
  \end{subfigure}
  \begin{subfigure}[b]{0.14\linewidth}
    \centering
    \includegraphics[width=\linewidth]{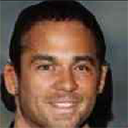}
    \captionsetup{justification=centering}
    \caption{JPEG\\Quality 35\%\\0.75 bit acc}
  \end{subfigure}
  \begin{subfigure}[b]{0.14\linewidth}
    \centering
    \includegraphics[width=\linewidth]{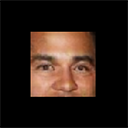} 
    \captionsetup{justification=centering}
    \caption{Cropping\\Crop size 64\\0.80 bit acc}
  \end{subfigure}
  \begin{subfigure}[b]{0.14\linewidth}
    \centering
    \includegraphics[width=\linewidth]{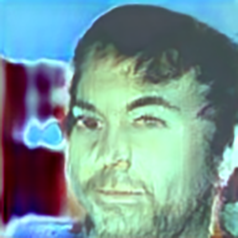} 
    \captionsetup{justification=centering}
    \caption{Quantize\\Precision $10^0$\\0.64 bit acc}
  \end{subfigure}
  \begin{subfigure}[b]{0.14\linewidth}
    \centering
    \includegraphics[width=\linewidth]{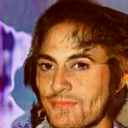} 
    \captionsetup{justification=centering}
    \caption{Model noise\\Std 0.16\\0.77 bit acc}
  \end{subfigure}
  \caption{Perturbed image samples from the fingerprinted ProGAN and the corresponding fingerprint detection accuracy. The detection still performs robustly (bitwise accuracy $\geq 0.75$) even when the image quality heavily deteriorates.}
  \label{fig:perturbation_samples} 
\end{figure*}

\textbf{Results.} We evaluate the artificial fingerprint detection over 50k images from a fingerprinted ProGAN. We plot the bitwise accuracy w.r.t. the amount of perturbations in Figure~\ref{fig:robutstness_plots} (see the supplementary material for additional results of ProGAN trained on LSUN \textit{Bedroom}). 
We observe:

(1) For all the image perturbations, fingerprint detection accuracy drops monotonously as we increase the amount of perturbation, while for small perturbations accuracy drops rather slowly. We consider accepting accuracy $\geq75\%$ as a threshold ($p$-value $=2.8\times10^{-7}$). This results in the working range w.r.t. each perturbation: Gaussian noise standard deviation $\sim[0.0, 0.05]$, Gaussian blur kernel size $\sim[0, 5]$, JPEG compression quality $\sim[50, 100]$, center cropping size $\sim[86, 128]$, quantization decimal precision $\leq10^{-1}$, and model noise standard deviation $\sim[0.0, 0.18]$, which are reasonably wide ranges in practice.

(2) For image perturbations (the left four subplots) outside the above working ranges, the reference upper bounds drop even faster and the margins to the testing curves shrink quickly, indicating that the detection deterioration is irrelevant to model training but rather relevant to the heavy quality deterioration of training images.

(3) For model perturbations (the right two subplots) outside the above working ranges, image quality deteriorates faster than fingerprint accuracy: even before the accuracy gets lower than $75\%$, FID has already increased by $>500\%$.

(4) As a result of (2) and (3), before fingerprint detection degenerates close to random guess ($\sim50\%$ accuracy), image quality has been heavily deteriorated by strong perturbations (Figure~\ref{fig:perturbation_samples}), which indicates that our fingerprints are more robust than image functionality itself in the case of these studied perturbations.

\textbf{Discussion on attacks.} Other attacks that require training counter models might be conceivable. For example, to train a model that removes the fingerprints from generated images (e.g. a denoising autoencoder). However, this would require attackers to have paired training images before and after the fingerprint embedding. In our scenario, we assume that the fingerprint encoder is not released which hinders this training data collection requirement. 

\subsection{Secrecy}
\label{sec:secrecy}

The presence of a fingerprint embedded in a generative model should not be easily detected by a third party, otherwise, it would be potentially manipulated. 

\textbf{Attack.} In order to quantify the secrecy, we formulate it as a binary classification problem: the presence of a fingerprint in an image. We follow the attack protocol in~\cite{zhu2018hidden} to perform the Artificial Training Sets (ATS) attack~\cite{lerch2016unsupervised}. We aim to classify between fingerprinted and non-fingerprinted test images
, but we have no supervision. The intuition is to train another steganography encoder to expand the test set and establish an artificial setting with known labels which enables supervised training. The original test class space is now a subspace of the artificial training class space and is separable by the training task.

The attack is as follows: We independently train another steganography encoder. We consider the original test images as negative training samples. Then, we apply the encoder twice to the test set to obtain extra images fingerprinted $2$ times (corresponding to originally non-fingerprinted images) or $3$ times (corresponding to originally fingerprinted images), which are regarded as positive training samples. Then we train an SVM classifier~\cite{chang2011libsvm} using such positive and negative samples, in order to separate between images fingerprinted $0$-$1$ time (original set), and the ones fingerprinted $2$-$3$ times (artificial training set). 

During testing, we first apply the encoder once to the test images so that the originally non-fingerprinted images are now fingerprinted $1$ time (belonging to $0$-$1$ class), and the originally fingerprinted images are now fingerprinted $2$ times (belonging to $2$-$3$ class). Then we can use the classifier to separate them and propagate the predictions back to the original images. We evaluate the attack on a set of $250$ fingerprinted deepfake images and $250$ non-fingerprinted deepfake images.

We investigate in three scenarios: (I) the weak attacker cannot access our $E$ and $D$, so he trains his shadow autoencoder and applies with arbitrary fingerprint codes; (II) the medium attacker can access our well-trained $E$ and $D$ but applies with arbitrary fingerprint codes; (III) the strong attacker can access our $E$ and $D$ and be partially aware of our fingerprint codes, so applies with the fingerprint codes only 10 bits different from ours.

\begin{table}[t!]
\center
\resizebox{0.9\linewidth}{!}{
\begin{tabular}{ccc|c}
\toprule
Attacker & Access to $E$ \& $D$ & Fgpt & Cls acc $\Downarrow$ \\
\midrule
Weak & No & Arbitrary & 0.503 \\
Medium & Yes & Arbitrary & 0.503 \\
Strong & Yes & 10-bit diff & 0.504 \\
\bottomrule
\end{tabular}}
\caption{Validation on the secrecy of fingerprints. The last column shows the binary classification accuracy of the presence of fingerprints, the smaller the more secret.}
\label{tab:secrecy_CNN_classifier}
\vspace{-4mm}
\end{table}

\textbf{Results.} In Table~\ref{tab:secrecy_CNN_classifier} all the attackers fail with the binary classification accuracy on the presence of fingerprint close to $0.5$ (random guess). 
\ning{It indicates our fingerprinting is secret enough from being detected by adversaries regardless of their access to our encoder and decoder.}

\subsection{Deepfake Detection}
\label{sec:detection}
In the previous sections, we showed that our fingerprinting solution is effective in transferring the fingerprints and meeting the other required criteria. 
We now discuss how to use it for deepfake detection and attribution. 

Unlike existing methods that detect intrinsic differences between the real and deepfake classes~\cite{ning2019iccv_gan_detection,zhang2019detecting,wang2020cnn,margret2020upconvolution}, we, \textit{standing for model inventors}, propose a proactive solution by embedding artificial fingerprints into generative models and consequently into the generated images. In practice, responsible model inventors, different from malicious deepfake users, should be \textit{eager}/\textit{willing} to do so. Then we convert the problem to verifying if one decoded fingerprint is in our fingerprint regulation database or not. 
Even with a non-perfect detection accuracy, we can still use our solution based on the null hypothesis test in Section~\ref{sec:method}. We consider deepfake verification given $\geq75\%$ ($p$-value $=2.8\times10^{-7}$) bit matching. This is feasible based on two assumptions: (1) The decoded fingerprint of a real image is random; and (2) the fingerprint capacity is large enough such that the random fingerprint from a real image is unlikely to collide with a regulated fingerprint in the database. The second condition is trivial to satisfy, considering we sample fingerprints $\mathbf{w} \in \{0,1\}^n$ and $n=100$. $2^{100}$ is a large enough capacity. Then we validate the first assumption by the deepfake detection experiments below. 

\begin{table}[t!]
\center
\resizebox{0.95\linewidth}{!}{%
\begin{tabular}{lll|cc}\toprule
 & & & Detection & Attribution \\
Dataset & Model & Detector & acc $\Uparrow$ & acc $\Uparrow$ \\
\midrule
\multirow{9}{*}{CelebA} & \multirow{3}{*}{ProGAN} & \cite{ning2019iccv_gan_detection} & 0.508 & 0.235\\
 & & \cite{wang2020cnn} & 0.924 & N/A\\
 & & Ours & 1.000 & 1.000\\
\cmidrule{2-5}
 & \multirow{3}{*}{StyleGAN} & \cite{ning2019iccv_gan_detection} & 0.497 & 0.168\\
 & & \cite{wang2020cnn} & 0.906 & N/A\\
 & & Ours & 1.000 & 1.000 \\
\cmidrule{2-5}
 & \multirow{3}{*}{StyleGAN2} & \cite{ning2019iccv_gan_detection} & 0.500 & 0.267\\
 & & \cite{wang2020cnn} & 0.895 & N/A\\
 & & Ours & 1.000 & 1.000 \\
\midrule
\multirow{9}{*}{LSUN} & \multirow{3}{*}{ProGAN} & \cite{ning2019iccv_gan_detection} & 0.493 & 0.597\\
  & & \cite{wang2020cnn} & 0.952 & N/A\\
  & & Ours & 1.000 & 1.000 \\
\cmidrule{2-5}
 & \multirow{3}{*}{StyleGAN} & \cite{ning2019iccv_gan_detection} & 0.499 & 0.366\\
 & & \cite{wang2020cnn} & 0.956 & N/A\\
 \textit{Bedroom} & & Ours & 1.000 & 1.000\\
\cmidrule{2-5}
 & \multirow{3}{*}{StyleGAN2} & \cite{ning2019iccv_gan_detection} & 0.491 & 0.267 \\
 & & \cite{wang2020cnn} & 0.930 & N/A\\
 & & Ours & 1.000 & 1.000 \\
\midrule
\multirow{6}{*}{LSUN} & \multirow{2}{*}{ProGAN} & \cite{wang2020cnn} & 0.951 & N/A\\
  & & Ours & 1.000 & 1.000 \\
\cmidrule{2-5}
 & \multirow{2}{*}{StyleGAN} & \cite{wang2020cnn} & 0.923 & N/A\\
\textit{Cat} & & Ours & 1.000 & 1.000\\
\cmidrule{2-5}
 & \multirow{2}{*}{StyleGAN2} & \cite{wang2020cnn} & 0.905 & N/A\\
 & & Ours & 1.000 & 1.000 \\
\midrule
\multirow{2}{*}{CIFAR-10} & \multirow{2}{*}{BigGAN} & \cite{wang2020cnn} & 0.815 & N/A\\
  & & Ours & 1.000 & 1.000 \\
\midrule
\multirow{2}{*}{\textit{Horse}$\rightarrow$\textit{Zebra}} & \multirow{2}{*}{CUT} & \cite{wang2020cnn} & 0.836 & N/A\\
  & & Ours & 1.000 & 1.000 \\
\midrule
\multirow{2}{*}{\textit{Cat}$\rightarrow$\textit{Dog}} & \multirow{2}{*}{CUT} & \cite{wang2020cnn} & 0.902 & N/A\\
  & & Ours & 1.000 & 1.000 \\
\bottomrule
\end{tabular}}
\caption{Deepfake detection and attribution accuracy ($\Uparrow$ indicates higher is better). \ning{\cite{wang2020cnn} is not applicable to the multi-source attribution scenarios in the last column.}}
\vspace{-3mm}
\label{tab:detection_attribution}
\end{table}

\textbf{Baselines.} 
\sahar{We compare to two recent state-of-the-art CNN-based deepfake detectors~\cite{ning2019iccv_gan_detection,wang2020cnn} as baselines.} \cite{ning2019iccv_gan_detection} is trained on 40k real images and 40k generated images equally from four generative models with distinct fingerprints.  
We consider the \textit{open-world} scenario where disjoint generative models are used in training and testing, to challenge the classifier's generalization. 
\ning{For \cite{wang2020cnn} we use the officially released model because they already claim improved generalization across different generation techniques.}

\textbf{Results.} \sahar{We compare our solution to the two baselines on a variety of generation applications, models, and datasets.} \ning{We test on 4k real images and 4k generated images equally from four generative models with distinct fingerprints.} We report deepfake detection accuracy in Table~\ref{tab:detection_attribution} fourth column. We observe:


(1) Our solution performs perfectly ($100\%$ accuracy) for all the cases, turning open-world deepfake detection into a trivial fingerprinting detection and matching problem.

(2) \cite{ning2019iccv_gan_detection} deteriorates to random guess ($\sim50\%$ accuracy) because of the curse of domain gap between training and testing models. In contrast, our solution benefits from being agnostic to generative models. It depends only on the presence of fingerprints rather than the discriminative cues that are overfitted during training.

(3) Our solution outperforms \cite{wang2020cnn} with clear margins. In particular, \cite{wang2020cnn} degenerates when model techniques evolve to be more powerful (from ProGAN to StyleGAN2), or condition on some input guidance. On the contrary, our proactive solution synergizes with this evolution with high fingerprint detection accuracy, and therefore, with perfect deepfake detection accuracy.

(4) In general, although \cite{wang2020cnn} generalizes better than \cite{ning2019iccv_gan_detection}, it is still subject to future adversarial evolution of generative models, which were witnessed rapidly progressing over the last few years. For example, \cite{wang2020cnn} was effectively evaded in~\cite{carlini2020evading} by extremely small perturbations. In contrast, our work offers higher sustainability in the long run by proactively enforcing a margin between real and generated images. \ning{This requires and enables responsible model inventors' disclosure against potential misuses of their models.}

\subsection{Deepfake Attribution}
\label{sec:attribution}

The goal of attribution is to trace the model source that generated a deepfake. It upgrades the binary classification in detection to multi-class classification. Our artificial fingerprint solution can be easily extended for attribution and enable us, standing for model inventors, to attribute responsibility to our users when misuses occur.



\textbf{Baseline.} \cite{wang2020cnn} is not applicable to multi-source attribution. We only compare to \cite{ning2019iccv_gan_detection} in the \textit{open-world} scenario, i.e., the training and testing sets of generative models are not fully overlapping. Given 40k generated images equally from four generative models with distinct fingerprints, we use \cite{ning2019iccv_gan_detection} to train four one-vs-all-the-others binary classifiers. During testing, all four classifiers are applied to an image. We assign the image to the class with the highest confidence if not all the classifiers reject that image. Otherwise, it is assigned to the unknown label.

\textbf{Results.} We compare our solution to \cite{ning2019iccv_gan_detection} on CelebA and LSUN \textit{Bedroom}. We test on 4k/4k generated images equally from four model sources that are in/out of the training set of \cite{ning2019iccv_gan_detection}. We report deepfake attribution accuracy in Table~\ref{tab:detection_attribution} last column. We obtain the same discoveries and conclusions as those of deepfake detection in Section~\ref{sec:detection}. The open-world attribution deteriorates for the CNN classifier~\cite{ning2019iccv_gan_detection} while our fingerprinting solution maintains the perfect ($100\%$) accuracy.

\section{Conclusion}
Detecting deepfakes is a complex problem due to the rapid development of generative models and the possible adversarial countermeasure techniques. For the sake of sustainability, we investigate a proactive solution on the model inventors' side to make deepfake detection agnostic to generative models. We root deepfake detection into training data, and demonstrate the transferability of artificial fingerprints from training data to a variety of generative models. Our empirical study shows several beneficial properties of fingerprints, including universality, fidelity, robustness, and secrecy. Experiments demonstrate our perfect detection and attribution accuracy that outperforms two recent state of the art. As there have been recent concerns about the release of powerful generative techniques, our solution closes the responsibility loop between publishing pre-trained generative model inventions and their possible misuses. It opens up possibilities for inventors' responsibility disclosure by allocating each model a unique fingerprint.

\section*{Acknowledgement}
Ning Yu was partially supported by Twitch Research Fellowship. Vladislav Skripniuk was partially supported by IMPRS scholarship from Max Planck Institute. This work was also supported, in part, by the DARPA SemaFor (HR001119S0085) program. Any opinions, findings, conclusions, or recommendations expressed in this material are those of the authors and do not necessarily reflect the views of the DARPA. We thank David Jacobs, Matthias Zwicker, Abhinav Shrivastava, Yaser Yacoob, and Apratim Bhattacharyya for constructive discussion and advice.

{\small
\bibliographystyle{ieee_fullname}
\bibliography{main}

\begin{thebibliography}{10}\itemsep=-1pt

\bibitem{steghide}
steghide, http://steghide.sourceforge.net.

\bibitem{outguess}
outguess, http://www.outguess.org/.

\bibitem{adi2018turning}
Yossi Adi, Carsten Baum, Moustapha Cisse, Benny Pinkas, and Joseph Keshet.
\newblock Turning your weakness into a strength: Watermarking deep neural
  networks by backdooring.
\newblock In {\em USENIX}, 2018.

\bibitem{baluja2017hiding}
Shumeet Baluja.
\newblock Hiding images in plain sight: Deep steganography.
\newblock In {\em NeurIPS}, 2017.

\bibitem{brock2018BigGAN}
Andrew Brock, Jeff Donahue, and Karen Simonyan.
\newblock Large scale gan training for high fidelity natural image synthesis.
\newblock In {\em ICLR}, 2018.

\bibitem{brown2020language}
Tom~B. Brown, Benjamin Mann, Nick Ryder, Melanie Subbiah, Jared Kaplan,
  Prafulla Dhariwal, Arvind Neelakantan, Pranav Shyam, Girish Sastry, Amanda
  Askell, et~al.
\newblock Language models are few-shot learners.
\newblock In {\em arXiv}, 2020.

\bibitem{carlini2020evading}
Nicholas Carlini and Hany Farid.
\newblock Evading deepfake-image detectors with white-and black-box attacks.
\newblock In {\em CVPR Workshops}, 2020.

\bibitem{cayre2005watermarking}
Francois Cayre, Caroline Fontaine, and Teddy Furon.
\newblock Watermarking security: theory and practice.
\newblock In {\em TSP}, 2005.

\bibitem{chang2011libsvm}
Chih-Chung Chang and Chih-Jen Lin.
\newblock Libsvm: A library for support vector machines.
\newblock In {\em TIST}, 2011.

\bibitem{chen2019deepmarks}
Huili Chen, Bita~Darvish Rouhani, Cheng Fu, Jishen Zhao, and Farinaz
  Koushanfar.
\newblock Deepmarks: A secure fingerprinting framework for digital rights
  management of deep learning models.
\newblock In {\em ICMR}, 2019.

\bibitem{choi2020stargan}
Yunjey Choi, Youngjung Uh, Jaejun Yoo, and Jung-Woo Ha.
\newblock Stargan v2: Diverse image synthesis for multiple domains.
\newblock In {\em CVPR}, 2020.

\bibitem{cox2002digital}
Ingemar Cox, Matthew Miller, Jeffrey Bloom, and Chris Honsinger.
\newblock {\em Digital watermarking}.
\newblock Springer, 2002.

\bibitem{margret2020upconvolution}
Ricard Durall, Margret Keuper, and Janis Keuper.
\newblock Watch your up-convolution: Cnn based generative deep neural networks
  are failing to reproduce spectral distributions.
\newblock In {\em CVPR}, 2020.

\bibitem{durall2019unmasking}
Ricard Durall, Margret Keuper, Franz-Josef Pfreundt, and Janis Keuper.
\newblock Unmasking deepfakes with simple features.
\newblock {\em arXiv}, 2019.

\bibitem{frank2020dct2d_detect}
Joel Frank, Thorsten Eisenhofer, Lea Sch{\"o}nherr, Asja Fischer, Dorothea
  Kolossa, and Thorsten Holz.
\newblock Leveraging frequency analysis for deep fake image recognition.
\newblock In {\em ICML}, 2020.

\bibitem{fridrich2009steganography}
Jessica Fridrich.
\newblock {\em Steganography in digital media: principles, algorithms, and
  applications}.
\newblock Cambridge University Press, 2009.

\bibitem{gan14nips}
Ian Goodfellow, Jean Pouget-Abadie, Mehdi Mirza, Bing Xu, David Warde-Farley,
  Sherjil Ozair, Aaron Courville, and Yoshua Bengio.
\newblock Generative adversarial nets.
\newblock In {\em NeurIPS}, 2014.

\bibitem{gulrajani2017wgan}
Ishaan Gulrajani, Faruk Ahmed, Martin Arjovsky, Vincent Dumoulin, and Aaron~C
  Courville.
\newblock Improved training of wasserstein gans.
\newblock In {\em NeurIPS}, 2017.

\bibitem{hayes2017generating}
Jamie Hayes and George Danezis.
\newblock Generating steganographic images via adversarial training.
\newblock In {\em NeurIPS}, 2017.

\bibitem{he2021beyond}
Yang He, Ning Yu, Margret Keuper, and Mario Fritz.
\newblock Beyond the spectrum: Detecting deepfakes via re-synthesis.
\newblock In {\em IJCAI}, 2021.

\bibitem{heusel2017gans}
Martin Heusel, Hubert Ramsauer, Thomas Unterthiner, Bernhard Nessler, and Sepp
  Hochreiter.
\newblock Gans trained by a two time-scale update rule converge to a local nash
  equilibrium.
\newblock In {\em NeurIPS}, 2017.

\bibitem{holub2012designing}
Vojt{\v{e}}ch Holub and Jessica Fridrich.
\newblock Designing steganographic distortion using directional filters.
\newblock In {\em WIFS}, 2012.

\bibitem{holub2014universal}
Vojt{\v{e}}ch Holub, Jessica Fridrich, and Tom{\'a}{\v{s}} Denemark.
\newblock Universal distortion function for steganography in an arbitrary
  domain.
\newblock In {\em EURASIP JIS}, 2014.

\bibitem{isola2017image}
Phillip Isola, Jun-Yan Zhu, Tinghui Zhou, and Alexei~A Efros.
\newblock Image-to-image translation with conditional adversarial networks.
\newblock In {\em CVPR}, 2017.

\bibitem{misinformationPoliticalCampaigns}
Charlotte Jee.
\newblock An indian politician is using deepfake technology to win new voters.
\newblock 2020.

\bibitem{karras2017ProGAN}
Tero Karras, Timo Aila, Samuli Laine, and Jaakko Lehtinen.
\newblock Progressive growing of gans for improved quality, stability, and
  variation.
\newblock In {\em ICLR}, 2018.

\bibitem{karras2019StyleGAN}
Tero Karras, Samuli Laine, and Timo Aila.
\newblock A style-based generator architecture for generative adversarial
  networks.
\newblock In {\em CVPR}, 2019.

\bibitem{karras2019StyleGAN2}
Tero Karras, Samuli Laine, Miika Aittala, Janne Hellsten, Jaakko Lehtinen, and
  Timo Aila.
\newblock Analyzing and improving the image quality of stylegan.
\newblock In {\em CVPR}, 2020.

\bibitem{krizhevsky2009learning}
Alex Krizhevsky, Geoffrey Hinton, et~al.
\newblock Learning multiple layers of features from tiny images.
\newblock Technical report, 2009.

\bibitem{ledig2017photo}
Christian Ledig, Lucas Theis, Ferenc Husz{\'a}r, Jose Caballero, Andrew
  Cunningham, Alejandro Acosta, Andrew Aitken, Alykhan Tejani, Johannes Totz,
  Zehan Wang, et~al.
\newblock Photo-realistic single image super-resolution using a generative
  adversarial network.
\newblock In {\em CVPR}, 2017.

\bibitem{lerch2016unsupervised}
Daniel Lerch-Hostalot and David Meg{\'\i}as.
\newblock Unsupervised steganalysis based on artificial training sets.
\newblock In {\em EAAI}, 2016.

\bibitem{liu2015faceattributes}
Ziwei Liu, Ping Luo, Xiaogang Wang, and Xiaoou Tang.
\newblock Deep learning face attributes in the wild.
\newblock In {\em ICCV}, 2015.

\bibitem{liu2020texture_fake}
Zhengzhe Liu, Xiaojuan Qi, Jiaya Jia, and Philip Torr.
\newblock Global texture enhancement for fake face detection in the wild.
\newblock In {\em CoRR}, 2020.

\bibitem{luo2020distortion}
Xiyang Luo, Ruohan Zhan, Huiwen Chang, Feng Yang, and Peyman Milanfar.
\newblock Distortion agnostic deep watermarking.
\newblock In {\em CVPR}, 2020.

\bibitem{marra2019gans}
Francesco Marra, Diego Gragnaniello, Luisa Verdoliva, and Giovanni Poggi.
\newblock Do gans leave artificial fingerprints?
\newblock In {\em MIPR}, 2019.

\bibitem{miyato2018spectral}
Takeru Miyato, Toshiki Kataoka, Masanori Koyama, and Yuichi Yoshida.
\newblock Spectral normalization for generative adversarial networks.
\newblock In {\em ICLR}, 2018.

\bibitem{park2020cut}
Taesung Park, Alexei~A. Efros, Richard Zhang, and Jun-Yan Zhu.
\newblock Contrastive learning for unpaired image-to-image translation.
\newblock In {\em ECCV}, 2020.

\bibitem{park2019semantic}
Taesung Park, Ming-Yu Liu, Ting-Chun Wang, and Jun-Yan Zhu.
\newblock Semantic image synthesis with spatially-adaptive normalization.
\newblock In {\em CVPR}, 2019.

\bibitem{pevny2010using}
Tom{\'a}{\v{s}} Pevn{\`y}, Tom{\'a}{\v{s}} Filler, and Patrick Bas.
\newblock Using high-dimensional image models to perform highly undetectable
  steganography.
\newblock In {\em IWIH}, 2010.

\bibitem{radford2015dcgan}
Alec Radford, Luke Metz, and Soumith Chintala.
\newblock Unsupervised representation learning with deep convolutional
  generative adversarial networks.
\newblock In {\em ICLR}, 2016.

\bibitem{radford2019language}
Alec Radford, Jeff Wu, Rewon Child, David Luan, Dario Amodei, and Ilya
  Sutskever.
\newblock Language models are unsupervised multitask learners.
\newblock In {\em arXiv}, 2019.

\bibitem{fakejournalism2}
Dan Robitzski.
\newblock Someone used deepfake tech to invent a fake journalist.
\newblock 2020.

\bibitem{ronneberger2015u}
Olaf Ronneberger, Philipp Fischer, and Thomas Brox.
\newblock U-net: Convolutional networks for biomedical image segmentation.
\newblock In {\em MICCAI}, 2015.

\bibitem{rouhani2019deepsigns}
Bita~Darvish Rouhani, Huili Chen, and Farinaz Koushanfar.
\newblock Deepsigns: an end-to-end watermarking framework for protecting the
  ownership of deep neural networks.
\newblock In {\em ASPLOS}, 2019.

\bibitem{tancik2019stegastamp}
Matthew Tancik, Ben Mildenhall, and Ren Ng.
\newblock Stegastamp: Invisible hyperlinks in physical photographs.
\newblock In {\em CVPR}, 2020.

\bibitem{thies2016face2face}
Justus Thies, Michael Zollhofer, Marc Stamminger, Christian Theobalt, and
  Matthias Nie{\ss}ner.
\newblock Face2face: Real-time face capture and reenactment of rgb videos.
\newblock In {\em CVPR}, 2016.

\bibitem{uchida2017embedding}
Yusuke Uchida, Yuki Nagai, Shigeyuki Sakazawa, and Shin'ichi Satoh.
\newblock Embedding watermarks into deep neural networks.
\newblock In {\em ICMR}, 2017.

\bibitem{fakejournalism1}
James Vincent.
\newblock An online propaganda campaign used ai-generated headshots to create
  fake journalists.
\newblock 2020.

\bibitem{vukotic2018deep}
Vedran Vukoti{\'c}, Vivien Chappelier, and Teddy Furon.
\newblock Are deep neural networks good for blind image watermarking?
\newblock In {\em WIFS}, 2018.

\bibitem{wang2020cnn}
Sheng-Yu Wang, Oliver Wang, Richard Zhang, Andrew Owens, and Alexei~A Efros.
\newblock Cnn-generated images are surprisingly easy to spot... for now.
\newblock In {\em CVPR}, 2020.

\bibitem{yu2015lsun}
Fisher Yu, Ari Seff, Yinda Zhang, Shuran Song, Thomas Funkhouser, and Jianxiong
  Xiao.
\newblock Lsun: Construction of a large-scale image dataset using deep learning
  with humans in the loop.
\newblock {\em arXiv}, 2015.

\bibitem{yu2018generative}
Jiahui Yu, Zhe Lin, Jimei Yang, Xiaohui Shen, Xin Lu, and Thomas~S Huang.
\newblock Generative image inpainting with contextual attention.
\newblock In {\em CVPR}, 2018.

\bibitem{ning2019iccv_gan_detection}
Ning Yu, Larry~S Davis, and Mario Fritz.
\newblock Attributing fake images to gans: Learning and analyzing gan
  fingerprints.
\newblock In {\em ICCV}, 2019.

\bibitem{yu2020inclusive}
Ning Yu, Ke Li, Peng Zhou, Jitendra Malik, Larry Davis, and Mario Fritz.
\newblock Inclusive gan: Improving data and minority coverage in generative
  models.
\newblock In {\em ECCV}, 2020.

\bibitem{yu2021dual}
Ning Yu, Guilin Liu, Aysegul Dundar, Andrew Tao, Bryan Catanzaro, Larry Davis,
  and Mario Fritz.
\newblock Dual contrastive loss and attention for gans.
\newblock In {\em ICCV}, 2021.

\bibitem{yu2020responsible}
Ning Yu, Vladislav Skripniuk, Dingfan Chen, Larry Davis, and Mario Fritz.
\newblock Responsible disclosure of generative models using scalable
  fingerprinting.
\newblock {\em arXiv}, 2020.

\bibitem{zhang2020not}
Baiwu Zhang, Jin~Peng Zhou, Ilia Shumailov, and Nicolas Papernot.
\newblock Not my deepfake: Towards plausible deniability for machine-generated
  media.
\newblock {\em arXiv}, 2020.

\bibitem{zhang2018protecting}
Jialong Zhang, Zhongshu Gu, Jiyong Jang, Hui Wu, Marc~Ph Stoecklin, Heqing
  Huang, and Ian Molloy.
\newblock Protecting intellectual property of deep neural networks with
  watermarking.
\newblock In {\em CCS Asia}, 2018.

\bibitem{zhang2019invisible}
Ru Zhang, Shiqi Dong, and Jianyi Liu.
\newblock Invisible steganography via generative adversarial networks.
\newblock In {\em Multimedia Tools and Applications}, 2019.

\bibitem{zhang2019detecting}
Xu Zhang, Svebor Karaman, and Shih-Fu Chang.
\newblock Detecting and simulating artifacts in gan fake images.
\newblock In {\em WIFS}, 2019.

\bibitem{zhou2021deep}
Peng Zhou, Ning Yu, Zuxuan Wu, Larry~S Davis, Abhinav Shrivastava, and Ser-Nam
  Lim.
\newblock Deep video inpainting detection.
\newblock {\em arXiv}, 2021.

\bibitem{zhu2018hidden}
Jiren Zhu, Russell Kaplan, Justin Johnson, and Li Fei-Fei.
\newblock Hidden: Hiding data with deep networks.
\newblock In {\em ECCV}, 2018.

\bibitem{zhu2017unpaired}
Jun-Yan Zhu, Taesung Park, Phillip Isola, and Alexei~A Efros.
\newblock Unpaired image-to-image translation using cycle-consistent
  adversarial networks.
\newblock In {\em ICCV}, 2017.

\bibitem{zhu2017toward}
Jun-Yan Zhu, Richard Zhang, Deepak Pathak, Trevor Darrell, Alexei~A Efros,
  Oliver Wang, and Eli Shechtman.
\newblock Toward multimodal image-to-image translation.
\newblock In {\em NeurIPS}, 2017.

\end{thebibliography}
}

\clearpage
\section{Supplementary Material}
\renewcommand\thesubsection{\Alph{subsection}}
\subsection{Implementation Details}
\label{supp_sec:implementation}

\begin{figure*}[!t]
\centering
\includegraphics[width=\linewidth]{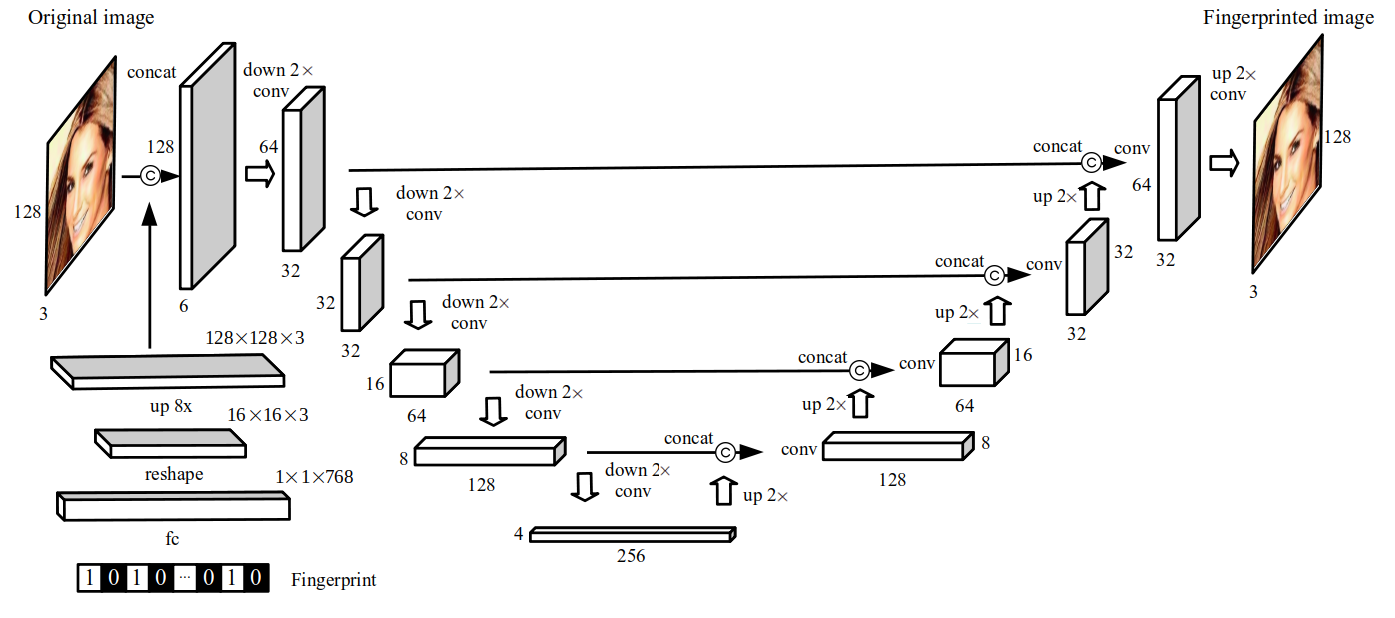}
\caption{Steganography encoder architecture.}
\label{supp_fig:encoder_architecture} 
\end{figure*}

\textbf{Steganography encoder.} The encoder is trained to embed a fingerprint into an image while minimizing the pixel difference between the input and stego images. We follow the technical details in~\cite{tancik2019stegastamp}. The binary fingerprint vector is first passed through a fully-connected layer and then reshaped as a tensor with one channel dimension and with the same spatial dimension of the cover image. We then concatenate this fingerprint tensor and the image along the channel dimension as the input to a U-Net architecture~\cite{ronneberger2015u}. The output of the encoder, the stego image, has the same size as that of the input image. Note that passing the fingerprint through a fully-connected layer allows for every bit of the binary sequence to be encoded over the entire spatial dimensions of the input image and flexible to the image size. The fingerprint length is set to $100$ as suggested in~\cite{tancik2019stegastamp}. The length of $100$ bits leads to a large enough space for fingerprint allocation while not having a side effect on the fidelity performance. We visualize an example of encoder architecture in Figure~\ref{supp_fig:encoder_architecture} with image size 128$\times$128 for CelebA and LSUN \textit{Bedroom}. For the other image sizes, the architectures are simply scaled up or down with more or fewer layers. 

\begin{figure*}[t!]
\centering
\includegraphics[width=\linewidth]{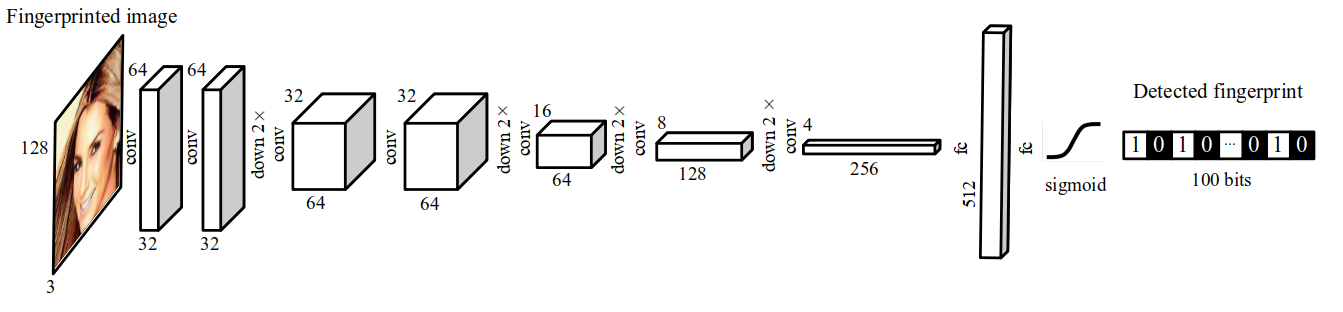}
\caption{Steganography decoder architecture.}
\label{supp_fig:decoder_architecture} 
\end{figure*}

\textbf{Steganography decoder.} The decoder is trained to detect the hidden fingerprint from the stego image. We follow the technical details in~\cite{tancik2019stegastamp}. It consists of a series of convolutional layers with kernel size $3$x$3$ and strides $\geq 1$, dense layers, and a sigmoid output activation to produce a final output with the same length as the binary fingerprint vector. We visualize an example of decoder architecture in Figure~\ref{supp_fig:decoder_architecture} with image size 128$\times$128 for CelebA and LSUN \textit{Bedroom}. For the other image sizes, the architectures are simply scaled up or down with more or fewer layers.

\textbf{Steganography training.} The encoder and decoder are jointly trained end-to-end w.r.t. the objective in Eq.~1 in the main paper and with randomly sampled fingerprints. The encoder is trained to balance fingerprint detection and image reconstruction. At the beginning of training, we set $\lambda=0$ to focus on fingerprint detection, otherwise, fingerprints cannot be accurately embedded into images. After the fingerprint detection accuracy achieves $95\%$ (that takes $3$-$5$ epochs), we increase $\lambda$ linearly up to $10$ within 3k iterations to shift our focus more on image reconstruction. We train the encoder and decoder for $30$ epochs in total. Given the batch size of $64$, it takes about 0.5/2/4 hours to jointly train a 32/128/256-resolution encoder and decoder using $1$ NVIDIA Tesla V100 GPU with 16GB memory.

Our steganography code is modified from the GitHub repository of StegaStamp~\cite{tancik2019stegastamp} official implementation\footnote{\url{https://github.com/tancik/StegaStamp}}. Our solution is favorably agnostic to generative model techniques because we only process the training data. Therefore, for generative model training, we directly refer to the corresponding GitHub repositories without any change: ProGAN~\cite{karras2017ProGAN}\footnote{\url{https://github.com/jeromerony/Progressive_Growing_of_GANs-PyTorch}}, StyleGAN~\cite{karras2019StyleGAN} and StyleGAN2~\cite{karras2019StyleGAN2} (config E)\footnote{\url{https://github.com/NVlabs/stylegan2}}, BigGAN~\cite{brock2018BigGAN}\footnote{\url{https://github.com/ajbrock/BigGAN-PyTorch}}, and CUT~\cite{park2020cut}\footnote{\url{https://github.com/taesungp/contrastive-unpaired-translation}}. 

\subsection{Additional Samples}

\begin{figure*}[t!]
\begin{subfigure}[b]{0.19\linewidth}
\centering
\includegraphics[width=\linewidth]{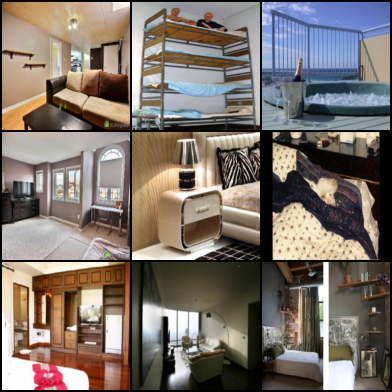}
\caption{}
\end{subfigure} 
\begin{subfigure}[b]{0.19\linewidth}
\centering
\includegraphics[width=\linewidth]{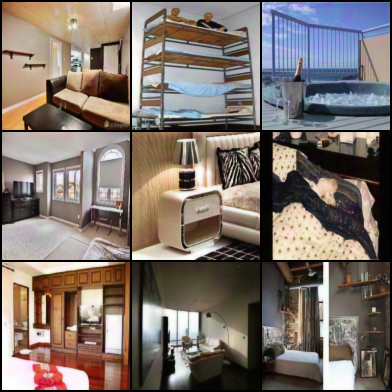} 
\caption{}
\end{subfigure} 
\begin{subfigure}[b]{0.19\linewidth}
\centering
\includegraphics[width=\linewidth]{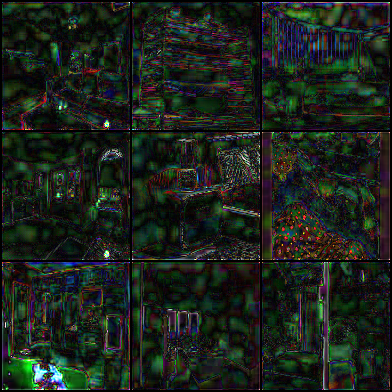}
\caption{}
\end{subfigure} 
\begin{subfigure}[b]{0.19\linewidth}
\centering
\includegraphics[width=\linewidth]{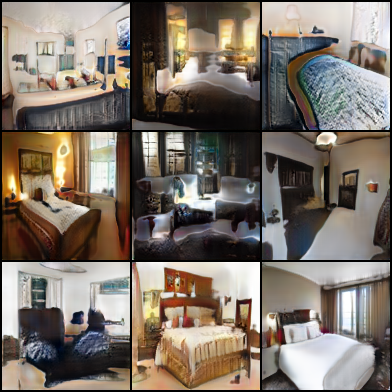}
\caption{}
\end{subfigure} 
\begin{subfigure}[b]{0.19\linewidth}
\centering
\includegraphics[width=\linewidth]{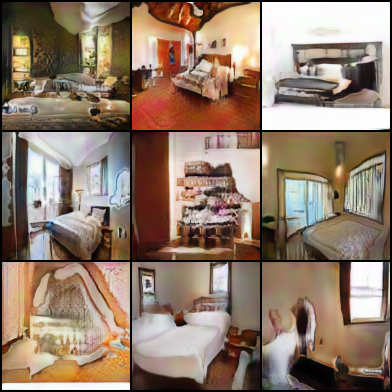}
\caption{}
\end{subfigure}
\caption{LSUN \textit{Bedroom} samples at 128$\times$128 for Table~\ref{tab:accuracy_fid} last two columns in the main paper, supplementary to Figure~\ref{fig:watermark_samples} in the main paper. (a) Original real training samples. (b) Fingerprinted real training samples. (c) The difference between (a) and (b), $10\times$ magnified for easier visualization. (d) Samples from the non-fingerprinted ProGAN. (e) Samples from the fingerprinted ProGAN.}
\label{supp_fig:watermark_samples_lsun_bedroom} 
\end{figure*}

\begin{figure*}[t!]
\begin{subfigure}[b]{0.48\linewidth}
\centering
\includegraphics[width=\linewidth]{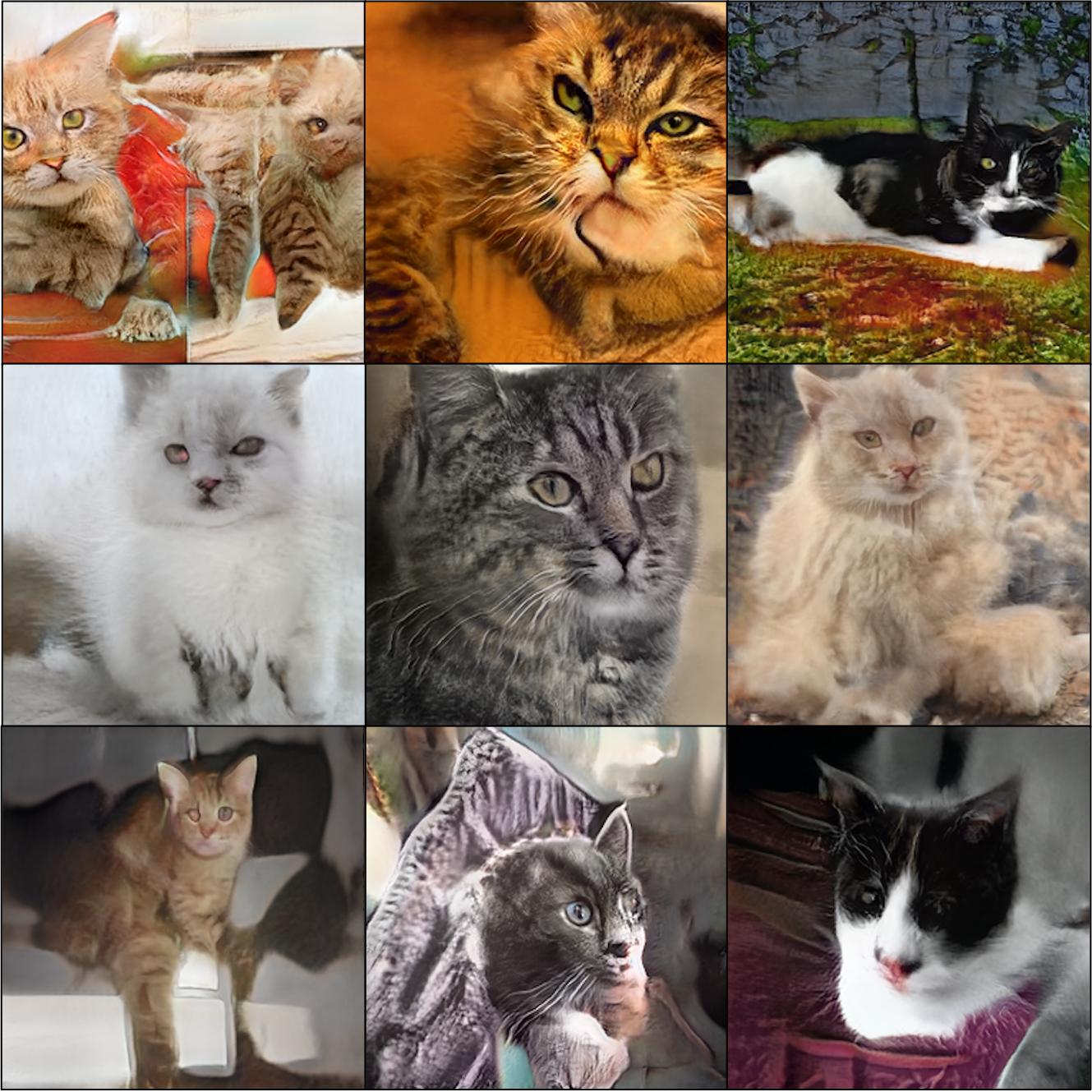}
\caption{}
\end{subfigure} 
\begin{subfigure}[b]{0.48\linewidth}
\centering
\includegraphics[width=\linewidth]{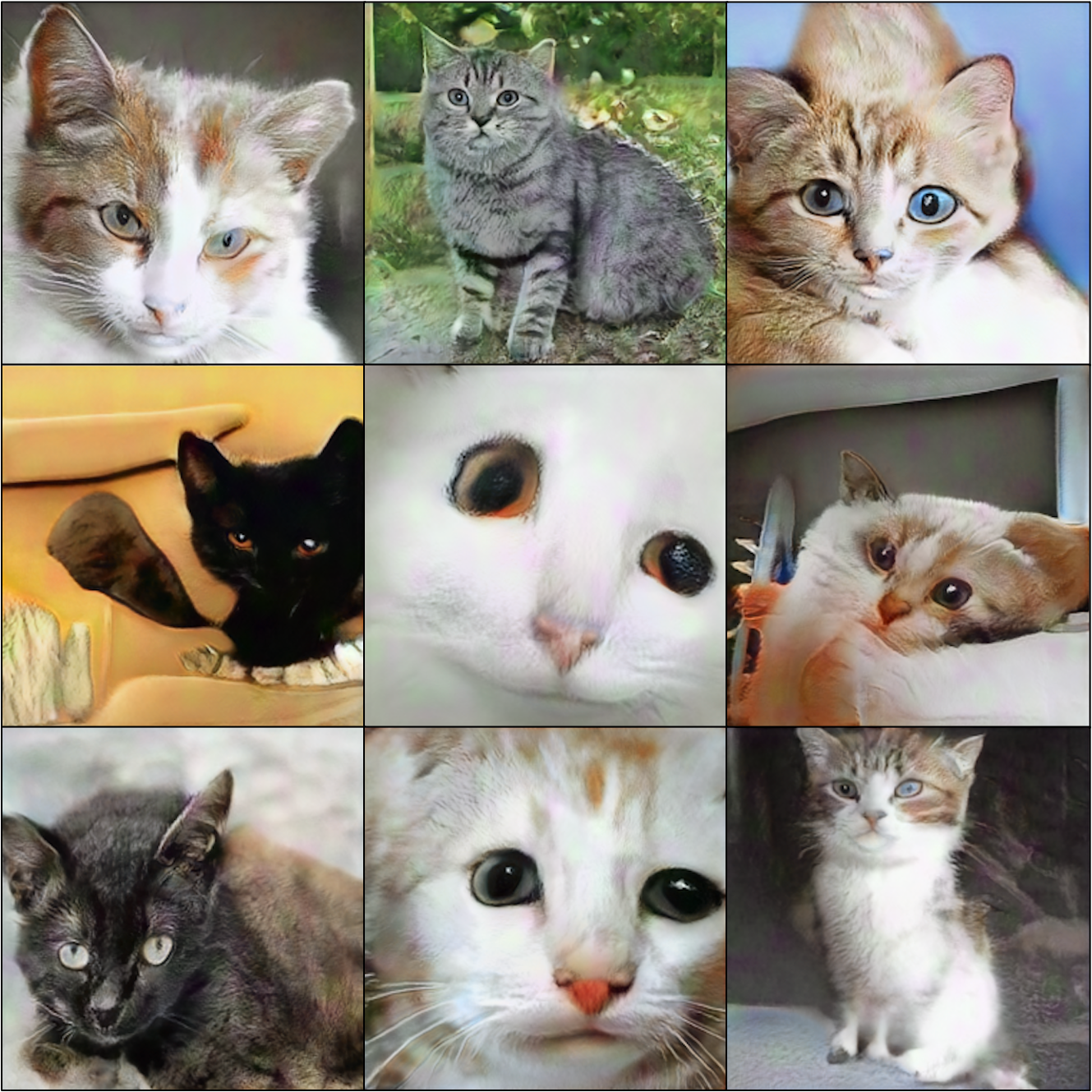} 
\caption{}
\end{subfigure} 
\caption{LSUN \textit{Cat} samples at 256$\times$256 for Table~\ref{tab:accuracy_fid} last two columns in the main paper, supplementary to Figure~\ref{fig:watermark_samples} in the main paper. (a) Samples from the non-fingerprinted StyleGAN2. (b) Samples from the fingerprinted StyleGAN2.}
\label{supp_fig:watermark_samples_lsun_cat} 
\end{figure*}

\begin{figure*}[t!]
\begin{subfigure}[b]{0.48\linewidth}
\centering
\includegraphics[width=\linewidth]{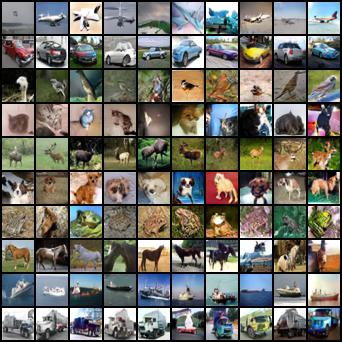}
\caption{}
\end{subfigure} 
\begin{subfigure}[b]{0.48\linewidth}
\centering
\includegraphics[width=\linewidth]{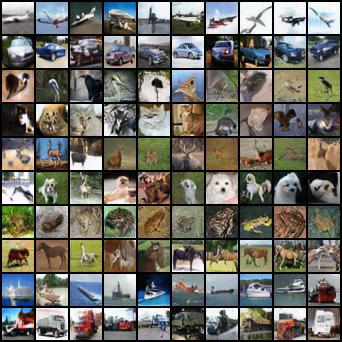} 
\caption{}
\end{subfigure} 
\caption{CIFAR-10 samples at 32$\times$32 for Table~\ref{tab:accuracy_fid} last two columns in the main paper, supplementary to Figure~\ref{fig:watermark_samples} in the main paper. (a) Samples from the non-fingerprinted BigGAN. (b) Samples from the fingerprinted BigGAN.}
\label{supp_fig:watermark_samples_cifar10} 
\end{figure*}

\begin{figure*}[t!]
\begin{subfigure}[b]{0.32\linewidth}
\centering
\includegraphics[width=\linewidth]{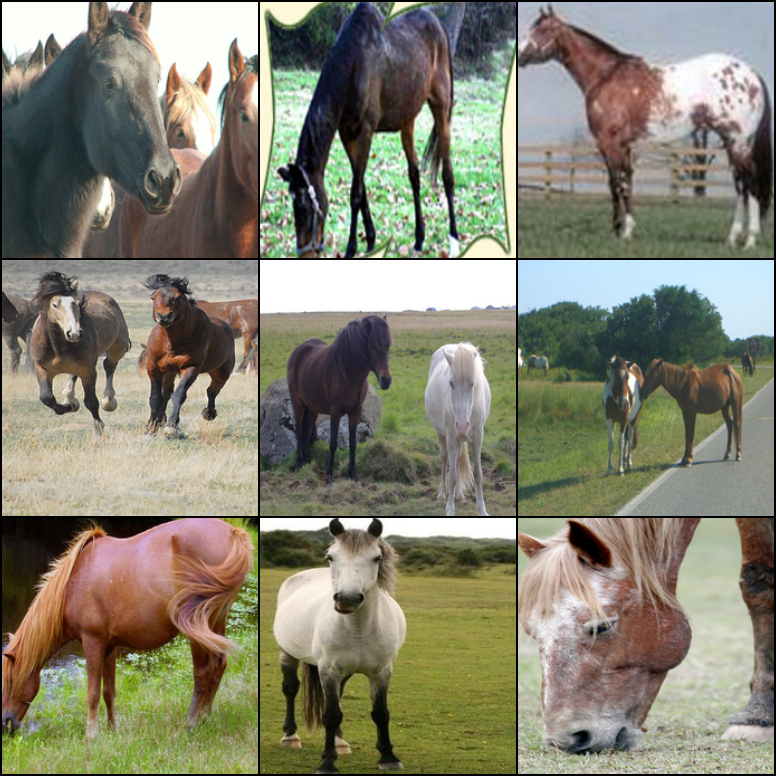}
\caption{}
\end{subfigure} 
\begin{subfigure}[b]{0.32\linewidth}
\centering
\includegraphics[width=\linewidth]{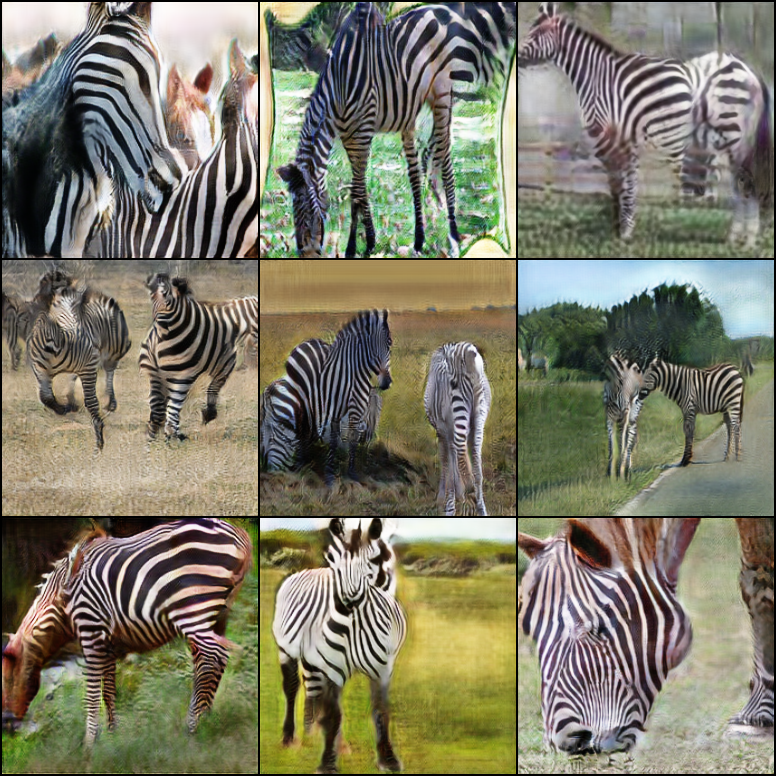}
\caption{}
\end{subfigure} 
\begin{subfigure}[b]{0.32\linewidth}
\centering
\includegraphics[width=\linewidth]{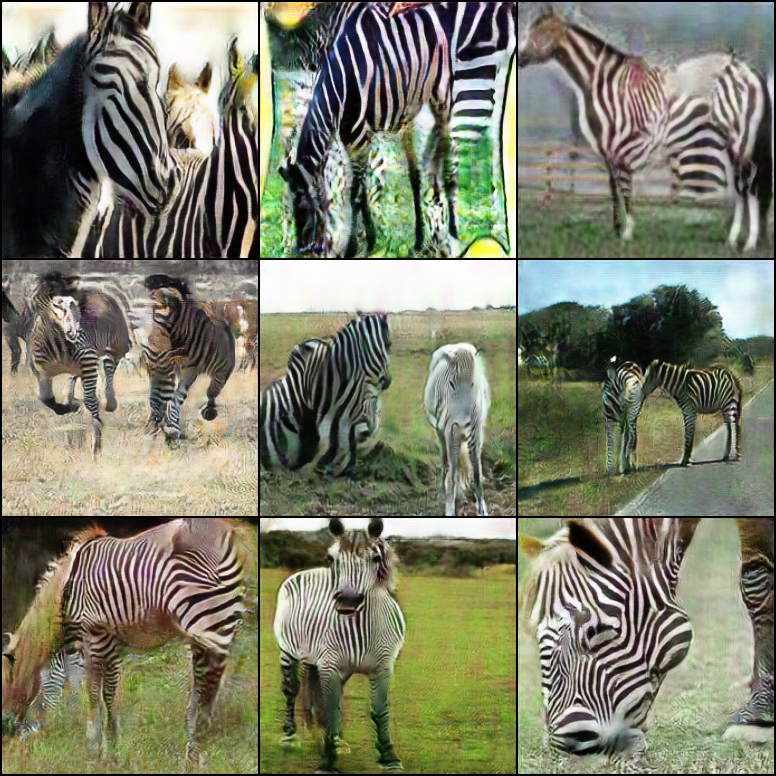} 
\caption{}
\end{subfigure} 
\caption{\textit{Horse}$\rightarrow$\textit{Zebra} samples at 256$\times$256 for Table~\ref{tab:accuracy_fid} last two columns in the main paper, supplementary to Figure~\ref{fig:watermark_samples} in the main paper. (a) Real source samples for input conditioning. (b) Samples from the non-fingerprinted CUT. (c) Samples from the fingerprinted CUT.}
\label{supp_fig:watermark_samples_horse2zebra} 
\end{figure*}

\begin{figure*}[t!]
\begin{subfigure}[b]{0.32\linewidth}
\centering
\includegraphics[width=\linewidth]{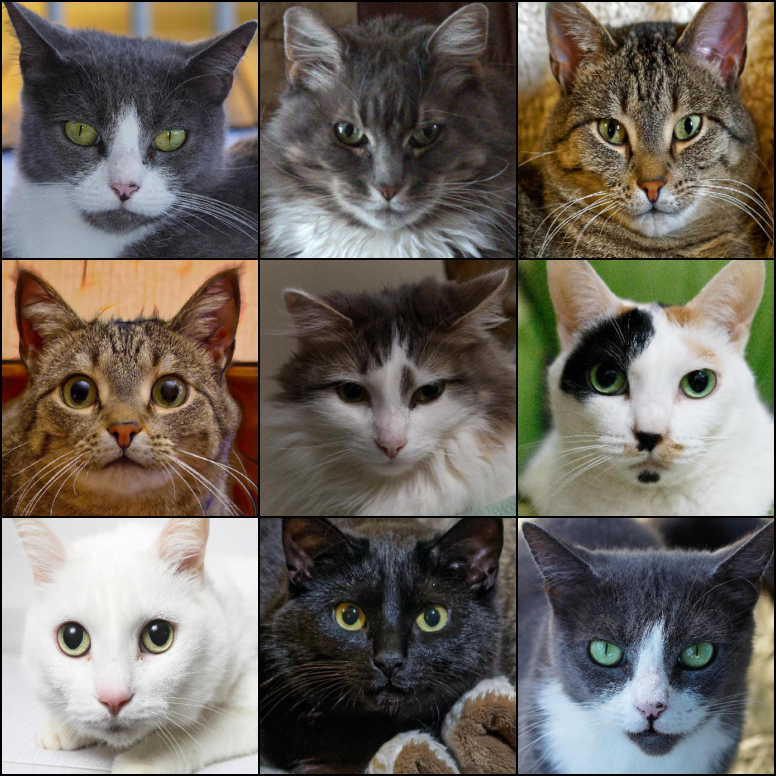}
\caption{}
\end{subfigure} 
\begin{subfigure}[b]{0.32\linewidth}
\centering
\includegraphics[width=\linewidth]{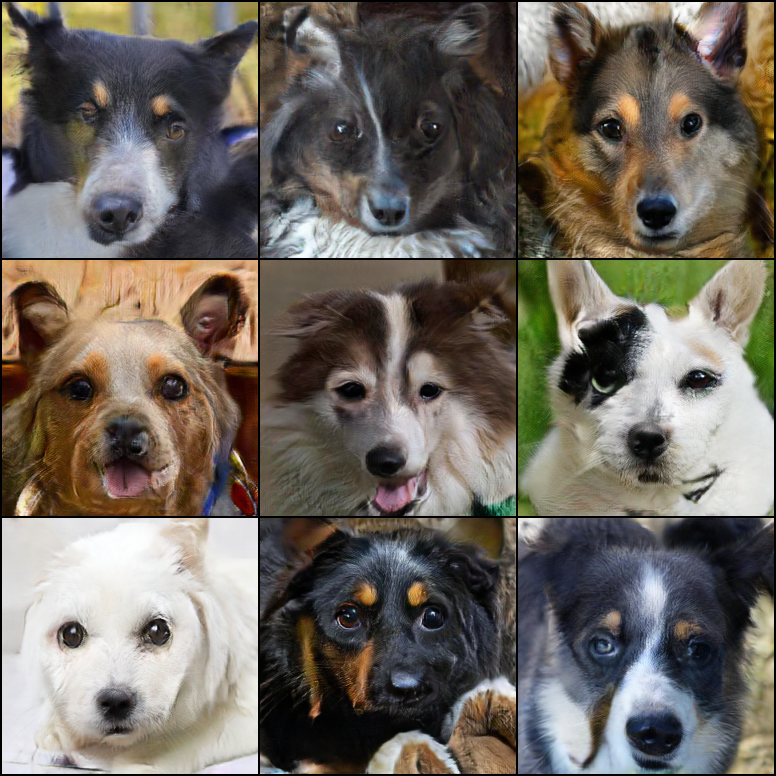}
\caption{}
\end{subfigure} 
\begin{subfigure}[b]{0.32\linewidth}
\centering
\includegraphics[width=\linewidth]{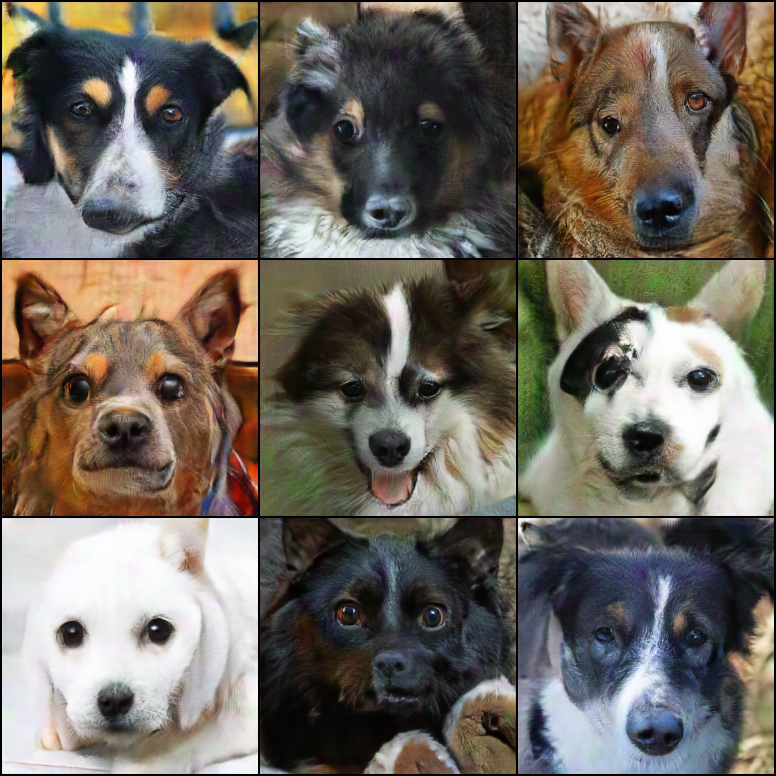} 
\caption{}
\end{subfigure} 
\caption{\textit{Cat}$\rightarrow$\textit{Dog} samples at 256$\times$256 for Table~\ref{tab:accuracy_fid} last two columns in the main paper, supplementary to Figure~\ref{fig:watermark_samples} in the main paper. (a) Real source samples for input conditioning. (b) Samples from the non-fingerprinted CUT. (c) Samples from the fingerprinted CUT.}
\label{supp_fig:watermark_samples_cat2dog} 
\end{figure*}

See Figure~\ref{supp_fig:watermark_samples_lsun_bedroom}, \ref{supp_fig:watermark_samples_lsun_cat}, \ref{supp_fig:watermark_samples_cifar10}, \ref{supp_fig:watermark_samples_horse2zebra}, and \ref{supp_fig:watermark_samples_cat2dog} for fingerprinted samples on a variety of generation applications, models, and datasets. We obtain the same conclusion as in Section~\ref{sec:fidelity} in the main paper: The fingerprints are imperceptibly transferred to the generative models and then to generated images.

\subsection{Robustness of ProGAN on LSUN \textit{Bedroom}}

\begin{figure*}[t!]
  \begin{subfigure}[b]{0.25\linewidth}
    \centering
    \includegraphics[width=\linewidth]{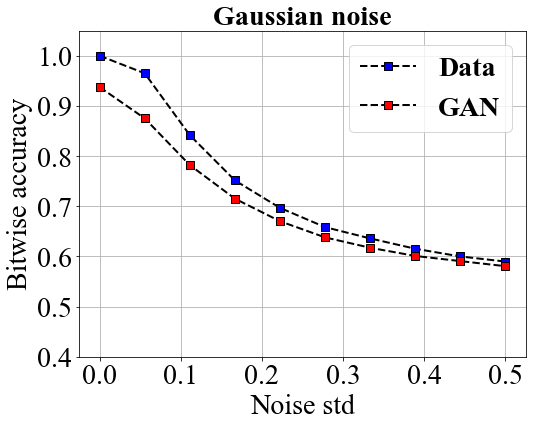}
  \end{subfigure}
  \begin{subfigure}[b]{0.25\linewidth}
    \centering
    \includegraphics[width=\linewidth]{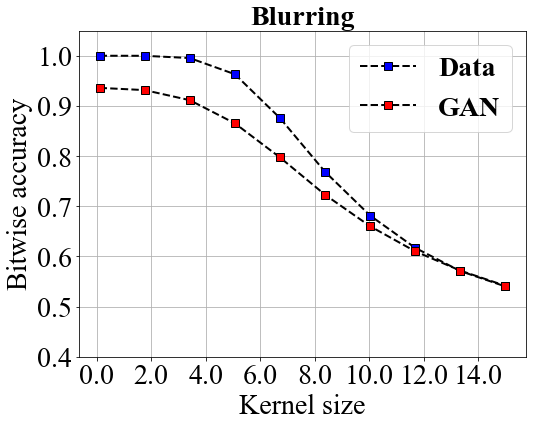} 
  \end{subfigure}
  \begin{subfigure}[b]{0.25\linewidth}
    \centering
    \includegraphics[width=\linewidth]{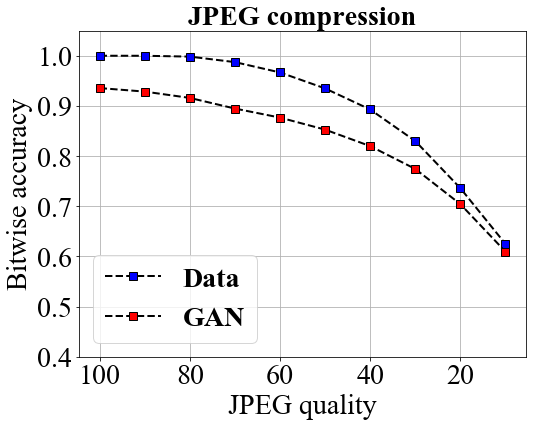} 
  \end{subfigure}
  \begin{subfigure}[b]{0.25\linewidth}
    \centering
    \includegraphics[width=\linewidth]{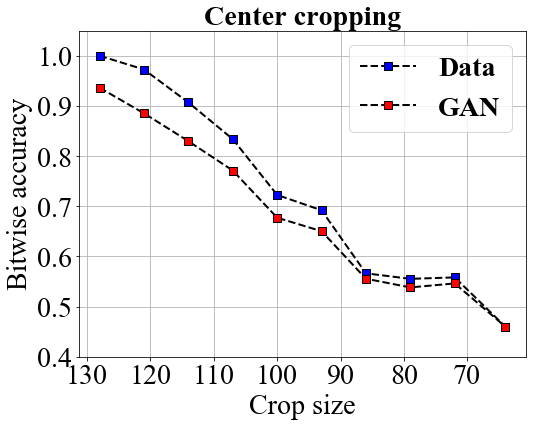} 
  \end{subfigure}
  \caption{
  Red plots show the artificial fingerprint detection in bitwise accuracy w.r.t. the amount of perturbations over ProGAN trained on LSUN \textit{Bedroom}. Blue dots represent detection accuracy on the fingerprinted real training images, which serve as the upper bound references for the red dots. This is supplementary to Figure~\ref{fig:robutstness_plots} in the main paper.}
 \label{supp_fig:robutstness_plots} 
\end{figure*}

We in additional experiment on the robustness of ProGAN on LSUN \textit{Bedroom}. We plot the bitwise accuracy w.r.t. the amount of perturbations in Figure~\ref{supp_fig:robutstness_plots}. We obtain the same conclusions as those in Section~\ref{sec:robustness} in the main paper. In specific, the working range w.r.t. each perturbation: Gaussian noise standard deviation $\sim[0.0, 0.1]$, Gaussian blur kernel size $\sim[0, 7]$, JPEG compression quality $\sim[30, 100]$, and center cropping size $\sim[108, 128]$, which are reasonably wide ranges in practice.

\end{document}